\documentclass{article}
\usepackage[english]{babel}
\usepackage{todonotes}
\usepackage{listings}
 \usepackage{color} 
 \definecolor{mygreen}{RGB}{28,172,0} 
 \definecolor{mylilas}{RGB}{170,55,241}
 \usepackage{graphics} 
 \usepackage{graphicx}
\usepackage{caption}
\usepackage{subcaption}
\usepackage{amsmath}
\usepackage{amssymb}
\usepackage{mathtools}
\usepackage{overpic}
\usepackage{amsthm}
\usepackage{changepage}
\usepackage{hyperref}

\usepackage{booktabs,array}

\usepackage{pgfplots}
\pgfplotsset{grid style={dotted,gray}}
\pgfplotsset{compat=newest}
\pgfplotsset{plot coordinates/math parser=false}
\newlength\figureheight
\newlength\figurewidth

\usepackage{stfloats} 
\usepackage{array, multirow}

\usepackage[pass]{geometry}
\usepackage[outdir=./]{epstopdf}
\usepackage{url}
\usepackage{tabstackengine}
\setstacktabbedgap{1.5ex}
\setstackgap{L}{1.2\normalbaselineskip}
 
\theoremstyle{definition}

\newtheorem{remark}{Remark}

\definecolor{KTHblue}{RGB}{25,105,188}
\definecolor{KTHlblue}{RGB}{22,159,219}
\definecolor{KTHyellow}{RGB}{251,186,0}
\definecolor{KTHred}{RGB}{176,9,48}
\definecolor{KTHlred}{RGB}{231,51,57}
\definecolor{KTHgreen}{RGB}{98,146,46}
\definecolor{KTHlgreen}{RGB}{175,202,11}
\definecolor{KTHpink}{RGB}{219,81,151}

\addto{\captionsenglish}{}
\renewcommand{\vec}[1]{\boldsymbol{#1}}
\newcommand{\vecmc}[1]{\boldsymbol{\mathcal{#1}}}

\addto{\captionsenglish}{}

\newgeometry{left=2.5cm,right =2.5cm,top= 2.5cm,bottom = 2.5cm}

\providecommand{\keywords}[1]{\textbf{Key words: } #1}

\usepackage[english]{nomencl}
\makenomenclature
\newcommand{\thickhline}{%
	\noalign {\ifnum 0=`}\fi \hrule height 1pt
	\futurelet \reserved@a \@xhline
}
\title{A Barrier Method for Contact Avoiding Particles in Stokes Flow}

\author{Anna Broms$^{1*}$ and Anna-Karin Tornberg$^{1}$\\\\
$^{1}$ Department of Mathematics\\
	KTH Royal Institute of Technology\\
	Lindstedtsv{\"a}gen 25, 114 28 Stockholm, Sweden\\
	$^*$e-mail: annabrom@kth.se, https://www.kth.se/profile/annabrom
}

\date{\today}

\begin{document}
\maketitle
\begin{abstract}
Rigid particles in a Stokesian fluid can physically not overlap, as a thin layer of fluid always separates a particle pair, exerting increasingly strong repulsive forces on the bodies for decreasing separations. Numerically, resolving these lubrication forces comes at an intractably large cost even for moderate system sizes. Hence, it can typically not be guaranteed that particle collisions and overlaps do not occur in a dynamic simulation, independently of the choice of method to solve the Stokes equations. In this work, non-overlap constraints, in terms of the Euclidean distance between boundary points on the particles, are represented via a barrier energy. We solve for the minimum magnitudes of repelling contact forces between any particle pair in contact to correct for overlaps by enforcing a zero barrier energy at the next time level, given a contact-free configuration at a previous instance in time. The method is tested using a multiblob method to solve the mobility problem in Stokes flow applied to suspensions of spheres, rods and boomerang shaped particles. Collision free configurations are obtained at all  instances in time. The effect of the contact forces on the collective order of a set of rods in a background flow that naturally promote particle interactions is also illustrated.

\end{abstract}
\textbf{Highlights}
\begin{itemize}
	 \itemsep0em
	 \item Strategy presented to avoid numerically introduced particle contacts in Stokes flow. 
	\item Non-colliding bodies guaranteed by enforcing zero barrier energy at each time-step.
	\item Contact force magnitudes are minimised for reduced effect on the physics of the system. 
	\item Contact forces and contact distances satisfy discrete complementarity. 
	 \item Both convex and non-convex rigid particles can be handled robustly.
\end{itemize}


\section{Introduction}\label{sec:intro}
We present an algorithm for contact forces between rigid particles immersed in an unbounded viscous fluid in 3D, introduced only when needed to avoid unphysical particle collisions and overlaps. Such collisions are caused from hard-to-avoid numerical artifacts and can also be the results of Brownian increments in a stochastic setting where thermal fluctuations in the fluid are considered. In this work, full hydrodynamic interaction is encountered for, meaning that there is a global coupling between all the particles in the system.

\begin{figure}[h!]
\centering
\begin{subfigure}[t]{0.23\textwidth}
	\centering
	\includegraphics[trim = {3.0cm 0.9cm 3cm 1cm},clip,width=1\textwidth]{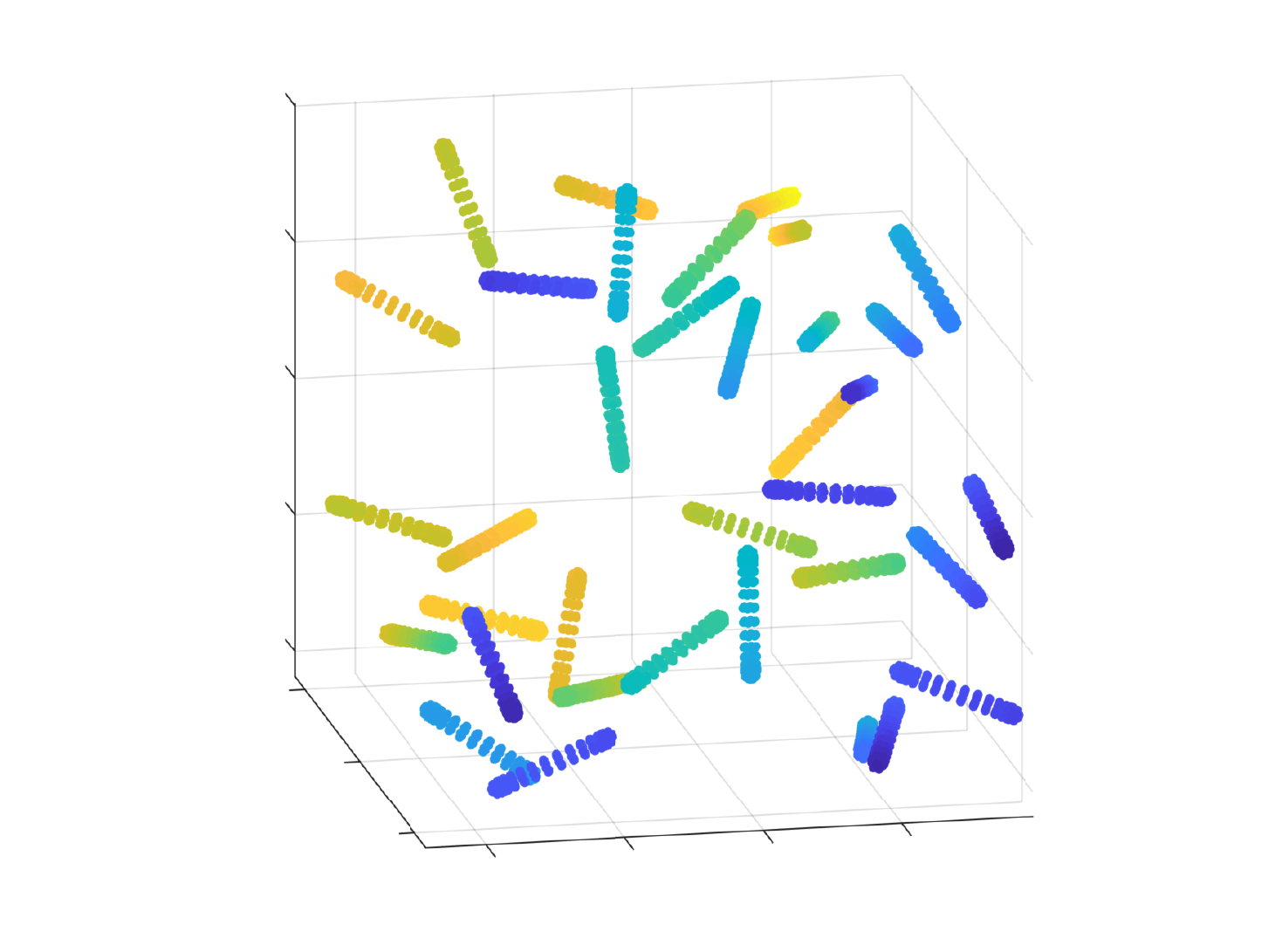}
	\caption{A suspension of slender rods.}
\end{subfigure}~~
\begin{subfigure}[t]{0.25\textwidth}
	\centering
	\includegraphics[trim = {3.5cm 0.9cm 3cm 1cm},clip,width=1\textwidth]{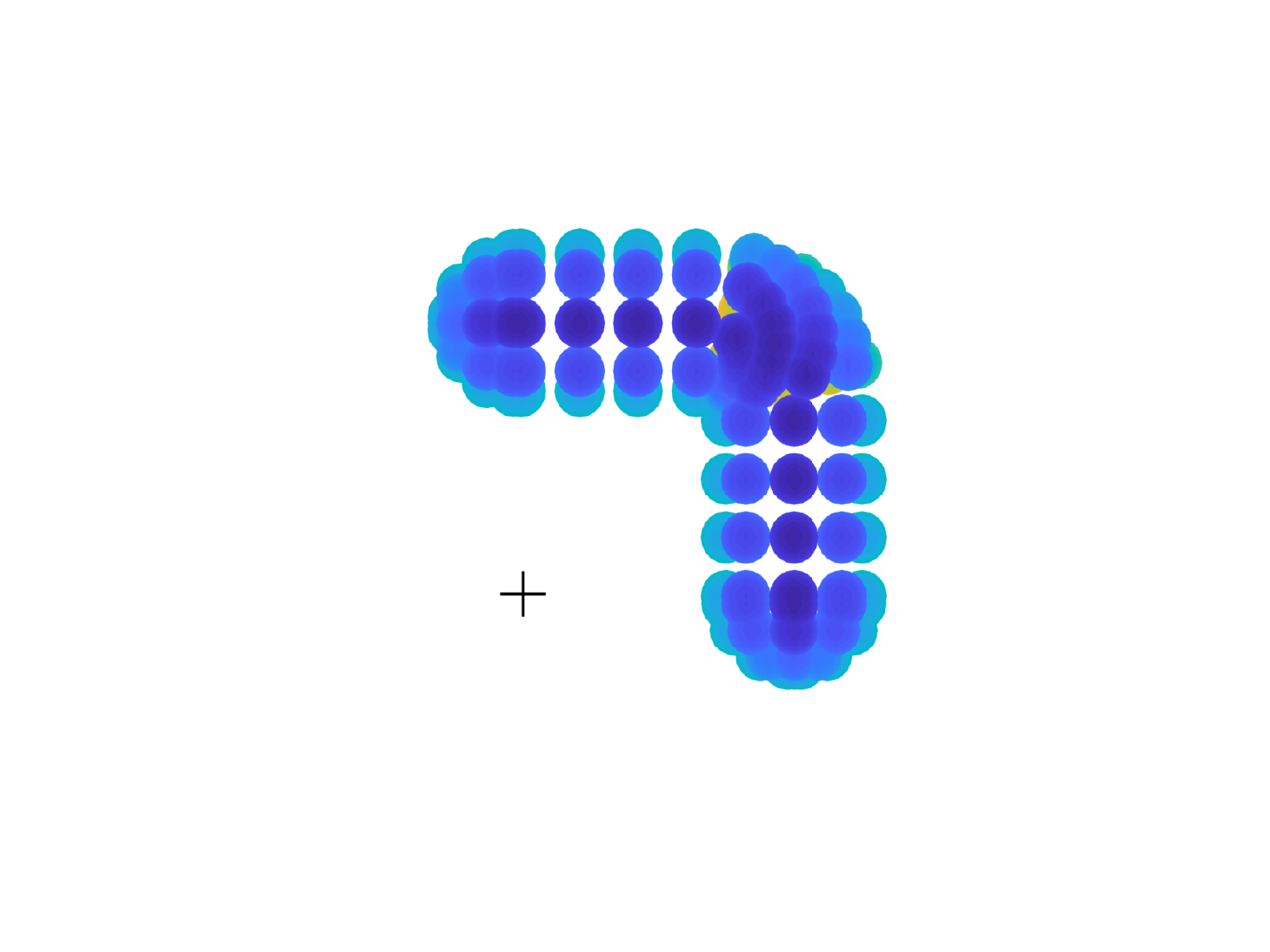}
	\caption{A multiblob boomerang with its center of mass indicated with a cross.}
	\label{boomerang}
\end{subfigure}
\caption{Multiblob particles in a Stokesian fluid. Colors indicate the depth in the figures.}
\label{multiblob_geom}
\end{figure}

We compute these hydrodynamic interactions with a cheap and fast technique: the so called rigid multiblob method, as carefully described in \cite{USABIAGA2016,Broms2022}. Note that this is only one of many numerical methods for Stokes flows and the contact avoiding strategy that we develop can be used also in combination with other techniques. The idea in the multiblob method is to model a rigid body by a collection of spheres or ``blobs'' and has been used in a very large number of works, see e.g.~\cite{Delong2015,Sprinkle2017,Brosseau2019,Brosseau2021,Fiore2019} and references therein. Each blob interacts hydrodynamically with all other blobs in the system in a pairwise manner and blobs belonging to the same particle are constrained to move as a rigid body via forces applied at each blob center, such that the blob forces sum to the net force and torque on the particle. Mathematically, one could view this as a regularised single layer boundary integral formulation, with the blob radius the regularisation parameter. Example multiblob geometries are displayed in Figure \ref{multiblob_geom}. For each instance in time, we solve the Stokes mobility problem, that is, given assigned external forces and torques on the particles, such as e.g.~gravity or some electrostatic forcing, the resulting particle translational and rotational velocities are computed. The inertia of the particles is typically negligible in the Stokes regime, and hence, the computed rigid body velocities can be used to update the particle positions, applying some suitable time-stepping scheme, and the configuration of particles can in this way be studied dynamically. By choosing the regularisation parameter and the surface where blobs are placed in relation to the true surface of the particle as the solution to a small off-line optimisation problem for each (axisymmetric) particle type, good accuracy in the particle velocities can be obtained for moderately separated particles even with coarse grids of the particle surfaces \cite{Broms2022}.

As particles get closer however, their interactions become increasingly difficult to accurately resolve. In the multiblob framework, accuracy  is suffering for particles close to being in contact, and this is a challenge independently of the choice of Stokes solver. Insufficient treatment of close interactions, caused either by numerical errors due to insufficient spatial resolution of the particle grid or by accumulated errors from the time-stepping scheme of the moving particles, can in the worst case lead to non-physical particle overlaps. Even if close interactions are resolved, and an adaptive time-stepping scheme is used, another unwanted effect is stalling when the adaptive time-step size becomes increasingly small as the time-step is adjusted to the stiff problem of particles in close proximity with increasingly strong repulsive forces. Lubrication forces between particles, which physically guarantees no-collision, can be resolved only using high fidelity methods with high spatial discretisation of the surfaces of the particles \cite{AfKlinteberg2014}. Both high temporal and spatial discretisation hence comes at a large cost, especially in dense suspensions. Keeping the surface grid resolution moderate, simulations become cheaper, but the accuracy is worsened and there is a large need for a contact avoiding strategy.  

Repelling contact forces should be introduced in such a way that the physical properties of the system are preserved. Contact forces are artificial to the system and introduced only to decrease the impact from the equally artificial numerical errors difficult to avoid for particles in close proximity. Hence, we want to find the smallest possible contact forces for particles to stay apart. At the same time, we also want to avoid introducing stiffness in the system, and eliminate the risk of having to take very small time-steps.  Such stiffness is often the draw back of applying a strongly repelling potential to avoid particle overlaps, as an alternative to contact resolution algorithms. Potentials of this type include Lennard-Jones or e.g.~a potential based on Hard Gaussian Overlap (HGO), which is designed for ellipsoids \cite{Berne1972} and discussed for rods in \cite{Varga2006}\footnote{For multiblob particles of a more general geometry, one could construct a potential from an HGO-potential centered on each individual blob building up the particle.}. Due to the problem with stiffness, it is difficult to guarantee non-overlapping particles with a potential-based method, and there is a risk that particles become ``soft'' \cite{Tao2005}.

Contact problems with particles \emph{not} immersed in a fluid has a richer literature \cite{Baraff1993,Anitescu1996,Tasora2008,Tasora2008a,Tur2009} with more benchmarks available. An overview of techniques for contact dynamic problems is found in \cite{Wriggers2006}. There is also a body of work in computer graphics \cite{Snyder1995,Harmon2011}, and of most relevance to this work, the recently introduced method of Incremental Potential Contact (IPC), which is a penalty method where colliding and overlapping configurations are penalised by a barrier energy \cite{Li2020,Ferguson2021,Li2021}. 

Different contact resolution techniques have been suggested to better resolve particle collisions in Stokes flow, see e.g.~the works \cite{Yamamoto1993,Yamamoto1995,Das2013,Delmotte2015b,Corona2017}, where no-slip boundary conditions are imposed at the point of contact, constraining the colliding particles to move with equal speed at the collision point. A complementarity formulation is favored in \cite{Lu2019,Yan2019,Yan2020,Bystricky2020,Yan2022,Kohl2021}, where the idea is to solve a nonlinear complementarity problem for contact force magnitudes and some definition of separation. The difficult-to-solve nonlinear problem is in turn approximated by one or a sequence of \emph{linear} complementarity problems (LCPs) that can be solved more easily. One method in this class is presented in the works by Yan et al., \cite{Yan2019,Yan2020,Yan2022}, and has the advantages that it is easily applicable to any hydrodynamic solver and is based on a geometric formulation that fulfills Newton's third law (forces on every pair of particles are balanced). One drawback, however, is that the method cannot guarantee non-overlapping configurations  at the end of each time-step, as the solution of a single LCP does not imply a solution to the original nonlinear problem. A second drawback is that non-convex particles cannot be handled. Another method of the same flavour for Stokesian fluids is based on so called Space-Time Interference Volumes (STIV) by Lu et al.~\cite{Lu2019,Lu2019a} and Bystricky et al.~\cite{Bystricky2020}. The STIV technique is inspired by the work of Harmon et al.~\cite{Harmon2011} in computer graphics. The general idea is to consider Stokes equation in variational form and after a candidate time-step, compute the volume in space-time swept out by the trajectories of the particles in contact. If the volume is negative, particles overlap. Enforcing this volume to be zero, by determining appropriate repulsion forces, gives rise to a complementarity problem. 
In an STIV approach, no-collision is obtained even for large time-steps by solving sequences of LCPs and particles that pass through each other during one time-step can be detected and the corresponding time-step corrected.  However, each STIV-formulation is strongly linked to a specific Stokes solver, the method is not easily applicable to general geometries, and in 3D, the STIV volume has to be efficiently computed in four dimensions \cite{Lu2019a,Lu2019}. Another difficulty is that using the STIV, contact forces are not automatically balanced and balanced forces are difficult to obtain (Newton's third law is \emph{not} satisfied) \cite{Bystricky2018t}.

We choose to approach the contact problem by introducing repelling contact forces to particles that come too close to each other in terms of the Euclidean distance. Key features of the method is that all contact forces are balanced via Newton's third law and that the next time-step in a sequence of time-steps is guaranteed to be contact free (in fact, by construction, a minimum separation distance is guaranteed between particles). The method is strongly inspired by the work of Yan et al. \cite{Yan2019,Yan2020,Yan2022} in how contact forces are geometrically motivated and by the work of Zorin and coauthors in the work on IPC, \cite{Li2020,Li2021,Ferguson2021}, in how non-overlapping constraints are set up. However, instead of penalising contact by a barrier energy as in \cite{Li2020,Li2021,Ferguson2021}, a barrier energy is rather used to represent a large number of non-overlap constraints and the complementarity condition with the associated contact force. In the work on IPC, \cite{Li2020,Li2021,Ferguson2021}, a tetrahedral or triangular mesh of the surface of each particle is considered, where distances are computed robustly between all pairs of edges and points to triangles respectively. In our work, particles are rigid and smooth and we can utilise the known parameterisation of the particle surfaces to compute distances robustly; given a set of points defining the grid of a particle surface, we instead flag the closest point of contact on the neighbouring particle to be part in the collision handling for each surface grid node within some set threshold of the other particle. This is possible as we are not dependent on second order derivatives of the distances in solving the optimisation problem with our formulation, in contrast to IPC; gradients of the distances with respect to particle coordinates suffice. The handling of the geometry and the constraints is also in contrast to the geometric approach by Yan et al, where only a single point of contact or "the most overlapping point" has to be determined, and more similar to the STIV technique, where multiple segments of the boundary can be flagged for the same particle contact pair. This is a beneficial property especially for non-convex particle geometries, or if surfaces are close to parallel, where the computed contact torque otherwise becomes very sensitive to the choice of contact point.   Details are outlined in Section \ref{measure}. Another difference in our work compared to \cite{Li2020,Li2021,Ferguson2021} is that particles are immersed in a Stokesian fluid, where intertia is negligible, and we solve for the force magnitudes rather than the particle coordinates in an implicit time-step as in IPC for rigid bodies \cite{Ferguson2021}. The dimension of the optimisation variable in the resulting optimisation problem hence becomes smaller. Keeping the rigidity of the particles is a non-issue (no additional constraints have to be enforced as discussed in \cite{Ferguson2021}). We utilise the linearity of the Stokes equations and formulate a minimisation problem for contact force magnitudes to solve in every time-step where contact occurs. Note that we do not minimise a combined, weighted, energy formulation as in IPC \cite{Li2020,Li2021,Ferguson2021}. Hence, there are no parameters to tune in our formulation for the importance of the collision constraints relative to the minimisation of other contributions to the total energy of the system.


The time-stepping schemes used both for STIV and the geometric approach by Yan et al.~is explicit (to avoid a large cost at every time-step) and of low order \cite{Lu2017,Yan2019,Yan2020,Yan2022}. The repulsion force is assumed to be constant over the course of one time-step and the basis for both types of contact algorithms in their vanilla version is explicit Euler. This is the time-stepping method that will be used for demonstration also in this work. An additional motivation to this choice is in settings where the contact resolution algorithm is coupled to Brownian motion, modelled by a stochastic differential equation, where it is difficult to obtain anything better than first order accuracy in time. Note however that there is nothing that prevents a higher order time-stepping scheme from being used if contact avoiding is applied in a deterministic setting. The most straight-forward idea is then to use a higher order time-stepping method for the particles not involved in any contacts. For particles to which repulsive forces are added in a certain time-step, there is no expected gain in using a high order method.

\subsection{The Stokes mobility problem}
Before introducing the optimisation problem, we start with some preliminaries. Each 3D particle in the fluid suspension can be described by its center coordinates $\vec{x}_i$ and rotation quaternion, $\vec q_i$. We collect these generalized coordinates for all the $N$ particles in the system in the vector $\vec{\mathcal{Q}}$ such that
\begin{equation}
\vec{\mathcal{Q}} = \left[
\vec{x}_1^T,  \vec q_1^T, \vec{x}_2^T, \vec q_2^T, \dots, \vec{x}_N^T, \vec q_N^T
\right]^T.
\end{equation}
Let $\vec{\mathcal U}$ be a vector of all rigid body velocities of the particles in the system, where $\vec u_i\in \mathbb R^3$ is the translational velocity and $\vec\omega_i\in \mathbb R^3$ the rotational velocity of particle $i$. Similarly, let $\vec{\mathcal F}_{\text{ext}}$ be a vector of all the externally applied forces $\vec f_i\in\mathbb R^3$ and torques $\vec t_i\in\mathbb R^3$ on the particles in the system: 
\begin{equation}
\vec{\mathcal U}=\begin{bmatrix}
\vec u_1^T& \vec\omega_1^T &\vec u_2^T &
\vec\omega_2^T &
\dots &
\vec u_{N}^T &
\vec\omega_N^T
\end{bmatrix}^T,\quad\vec{\mathcal F}_{\text{ext}} = 
\begin{bmatrix}
\vec f_1^T & \vec t_1^T & \vec f_2^T &\vec t_2^T & \dots & \vec f_N^T &\vec t_N^T
\end{bmatrix}^T.
\end{equation}
Now, the Stokes mobility problem can be stated as 
\begin{equation}\label{mobility}
\vec{\mathcal U} = \vecmc U_{\text{bg}}+\vecmc U_{\text{Brownian}}+\vec{\mathcal{M}}\vec{\mathcal F}_{\text{ext}},
\end{equation}
where $\vec{\mathcal{M}}$ is the mobility matrix of size $6N\times 6N$, with each $6\times 6$ block corresponding to the interaction between a specific pair of particles. The vector $\vecmc U_{\text{bg}}$ is the velocity contribution on the particles from an eventual background flow and $\vecmc U_{\text{Brownian}}$ is a vector of stochastic velocities included if thermal fluctuations are considered. Note that implementation-wise, $\vec{\mathcal M}$ is only to be interpreted symbolically and $\vec{\mathcal{M}}\vec{\mathcal F}_{\text{ext}}$ corresponds to solving Stokes equations along with no-slip boundary conditions using the multiblob method as described in \cite{Broms2022}, given forces and torques for all particles, stacked in $\vec{\mathcal F}_{\text{ext}}$.

With no contact forces present, $\vec{\mathcal{Q}}$ is related to the velocities in $\vecmc U$ as
\begin{equation}
\dot{\vec{\mathcal{Q}}} = \vec\Psi\vec{\mathcal U}, 
\end{equation}
with $\vec\Psi$ a geometry-dependent matrix relating velocities to particle positions and quaternions. 

When particles are in contact, the velocities are corrected using the action of a vector of contact forces $\vec{\mathcal F}_c$, such that 
 \begin{equation}\label{eq5}
 \dot{\vec{\mathcal{Q}}} = \vec\Psi\vec{\mathcal U}+\vec\Psi\vec{\mathcal{M}}\vec{\mathcal F}_c,
 \end{equation}
  with
 \begin{equation}\label{eq6}
 \vec{\mathcal F}_c = \vecmc{D}\vec{\lambda}.
 \end{equation}
The matrix $\vecmc D\in\mathbb R^{6N}\times\mathbb R^{N_c}$ determines the direction of contact forces and torques and depends on the particle configuration and geometries and the vector $\vec\lambda\in\mathbb R^{N_c}$ determines the contact force magnitudes, with $N_c$ the number of particle pairs in contact. Ideally, we would like to determine contact forces such that if contact forces are needed for particles to stay apart during the time-step, the force should be active and the corresponding force magnitude positive. On the other hand, if the particles stay separate without contact forces, the contact force magnitude should be zero and the contact force passive. The nature of the contact problem is hence a complementarity problem. It is also a nonlinear problem, as the distances at the next instance of time depends nonlinearly of the contact forces.

 \section{A barrier method for particles in contact}
 In practice, it is not advisable to let particles come so close to each other that they eventually touch, due to limitations in all numerical solvers for particles in Stokes flow -- the accuracy for almost touching particles is suffering if the resolution of the particles is not excessively high. For this reason, we define a contact to occur if the distance $d$ between points on particles in close proximity is smaller than a set buffer distance $\hat{d}$. The parameter $\hat{d}$ is typically set to ensure a prescribed accepted accuracy from the multiblob method\footnote{One could also consider to set $\hat{d}$ in relation to any inhomogenities on the surfaces of the physical particles modelled by our rigid particles.}. We would like to find contact force magnitudes $\vec\lambda$ such that this separation distance is guaranteed at the next time-step, i.e.
 \begin{equation}\label{overlap}
 d_i\left(\vecmc Q_{t+\Delta t}(\vec\lambda)\right)>\hat{d},\quad\text{for all $i$ pairs of points on distinct particles},
 \end{equation}
  with $d_i$ depending non-linearly on $\vecmc Q_{t+\Delta t}$, the particle configuration in the next time step, and the set of relevant $d_i$ for each contact pair properly defined in Section \ref{measure}. The complementarity condition for the particle distances and contact force magnitudes take the form 
   \begin{equation}\label{comp}
 \lambda_k\max\left(0,\min\limits_{i}\left\{d_i(\vecmc Q_{t+\Delta t})-\hat{d}\right\}\right)=0,\quad \text{$i$ associated with particle contact pair $k$ and }\lambda_k\geq 0.
 \end{equation}
 Instead of solving this (modified) nonlinear complementarity problem sharply, we will set up a contact optimisation problem to fulfill all constraints in \eqref{overlap} and still approximately solve \eqref{comp}. Note that depending on the number and concentration of particles in the system and their geometries, the number of constraints in \eqref{overlap} might be very large. This section explains how such a minimisation problem can be formulated and discuss modelling choices for the objective function, the constraints and the geometry of the contact forces. The constraints in \eqref{overlap} can be represented via a minimisation of a sum of indicator functions \cite{Boyd2009}:
  \begin{equation}\label{indicator}
  \min\limits_{\vec\lambda\geq \vec 0}\quad \sum_{i} I(d_i\left(\vecmc Q_{t+\Delta t}(\vec\lambda)\right)-\hat{d}),\quad\text{with } I(s) =\left\{\begin{aligned}
  0,\quad s\geq 0,\\
  \infty, \quad s<0.
  \end{aligned}\right.
  \end{equation}
 This representation is however difficult to work with and we therefore replace \eqref{indicator} with a smoother counterpart.
Introduce a so-called barrier function, $b$, that is zero for sufficiently large distances $d$ and increasingly large for distances smaller than the set threshold $\hat{d}$: 
 \begin{equation}\label{b_opt}
 b(d,\hat{d}) =  \left\{\begin{aligned}
 -(d-\hat{d})^2\ln \left(d/\hat{d}\right),\quad &0<d<\hat{d},\\
 0,\quad &d\geq \hat{d}.
 \end{aligned}\right. \\   
 \end{equation}
 We can collect all the non-overlapping constraints in \eqref{overlap} into what we define as a \emph{barrier energy}, mimicking the sum of indicators in \eqref{indicator}. A non-overlapping configuration is one with non-negative contact force magnitudes $\vec\lambda$ applied for $t\in[t,t+\Delta t]$ that minimises the barrier energy
 \begin{equation}\label{barren}
 \min\limits_{\vec\lambda\geq \vec 0}\quad\sum_ib\left( d_i(\vecmc Q_{t+\Delta t}),\hat{d}\right),
 \end{equation}
with the summation being over all geometries in contact (corresponding to all constraints in \eqref{overlap}). In the Stokesian fluid, the particle coordinates evolve with time according to 
 \begin{equation}
 \dot{\vecmc Q} = \vec\Psi\left(\vecmc U + \vecmc M\vecmc  D\vec\lambda\right).
 \end{equation}
 If we discretise this equation in time, we can express $\vecmc Q_{t+\Delta t}$  in terms of previous, known, coordinate vectors. With the forward Euler method, the minimisation problem becomes
 \begin{equation}\label{min_barrier}
 \begin{aligned}
  &\min\limits_{\vec\lambda\geq \vec 0}\quad&&\sum_ib\left( d_i(\vecmc Q_{t+\Delta t}),\hat{d}\right),\\
  &\text{s.t. } &&\vecmc{Q}_{t+\Delta t} = \vecmc Q_t+\Delta t\vec{\Psi}\left(\vecmc U + \vecmc M\vecmc  D\vec\lambda\right).
 \end{aligned}
 \end{equation}
 This is a constrained minimisation problem to solve for $\vec \lambda\in\mathbb R^{N_c}$, i.e.~with a scalar contact force magnitude per particle pair in contact. Note however that for some particle configurations, any  vector $\vec\lambda$ with sufficiently large magnitude would solve \eqref{min_barrier}. We hence need to constrain the solution by penalising large $\vec\lambda$, with a natural choice being 
 \begin{equation}\label{pot_barrier1}
 \min\limits_{\vec\lambda\geq \vec 0}\quad\|\vec\lambda\|+\alpha\sum\limits_ib\left( d_i\left(\vecmc Q_t+\Delta t\vec{\Psi}\left(\vecmc U + \vecmc M\vecmc  D\vec\lambda\right)\right),\hat{d}\right),
 \end{equation}
 where $\alpha$ is a parameter that we have to set to balance the minimisation of the barrier energy at the next time-step and the minimisation of contact forces.   The parameter $\alpha$ also has the role of enforcing the complementarity condition for the contact forces and contact distances (relative to the buffer region) in \eqref{comp}. A different way of saying the same thing is that we would like to pick $\lambda_k$ sufficiently small so that non-overlap is obtained, but where the force is active, the minimum contact distance should not be larger than $\hat{d}$ after applying the contact forces.  
 
For the choice of norm in \eqref{pot_barrier1}, we have two options: A 1-norm penalty of the contact force magnitudes, $\sum_i\lambda_i$, penalises the forces at all contacts to an equal extent and not only those where the magnitude is large. One could also choose the norm $\left(\vec\lambda^T\vecmc D^T\vecmc M \vecmc D\vec\lambda\right)^{1/2}$, or its square, corresponding to the dissipated energy induced by the contact forces. As a main goal is to minimise the impact of the artificial repulsive forces on the system, the dissipated energy due to the introduced contact forces could be a reasonable choice of objective function. For our Stokes solver, the rigid multiblob method, it is a known problem that the accuracy in $\vecmc M$ is suffering a lot for closely interacting particles \cite{Broms2022}. Hence, $\vec\lambda\vecmc D^T\vecmc M\vecmc D\vec\lambda$ is a very crude approximation to the true dissipated energy and a large error is associated with this quantity. For this reason we will use $\sum_i\lambda_i$ as our objective function of choice. See section \ref{hyper} for a numerical comparison of the two choices.
 
 Note that it is not trivial to find the best choice of $\alpha$ in \eqref{pot_barrier1} for general applicability as the magnitude of both $\vec\lambda$ and the barrier energy depends on the number of pairs of particles in contact and how much overlap a non-corrected time-step would yield. A suitable level of $\alpha$ also depends on the tolerance chosen as stopping criteria when solving the problem in \eqref{pot_barrier1}. This is a similar problem as the one encountered when minimising the energy formulation in the work on IPC, where a penalty has to be chosen in an adaptive manner to give the barrier energy the proper weight, see the supplementary in \cite{Li2020}. Here, we would like to avoid such hyperparameter tuning.  What \emph{is} known, however, is that the barrier energy is identically zero for a non-overlapping configuration. An alternative  formulation of the minimisation problem in \eqref{pot_barrier1} is therefore
 \begin{equation}\label{minnorm}
 \begin{aligned}
 &\min_{\vec\lambda\geq\vec 0}\quad\sum_i\lambda_i,\\
 &\text{s.t. } \sum\limits_ib\left( d_i\left(\vecmc Q_t+\Delta t\vec{\Psi}\left(\vecmc U + \vecmc M\vecmc  D\vec\lambda\right)\right),\hat{d}\right) = 0,
 \end{aligned}
 \end{equation}
 where contact force magnitudes are minimised while enforcing a zero barrier energy. The formulation in \eqref{minnorm} is simpler than \eqref{pot_barrier1} in the sense that one hyperparameter is reduced, but possibly more challenging as a nonlinear equality constraint has been added. Note that the Lagrangians of \eqref{pot_barrier1} and \eqref{minnorm} are identical, but in the case of \eqref{minnorm}, the parameter $\alpha$ is a Lagrange multiplier to be solved for instead of a set parameter. In the remaining of this work, we will consider the formulation in \eqref{minnorm} for determining contact forces.

\begin{remark}
	In contrast to in the work on IPC by Zorin et al., the value of the barrier energy itself is not used explicitly in the formulation presented in this paper but necessary for
	\begin{enumerate}
	\item Imposing a large number of constraints of minimum distances, as we seek only the force magnitudes where the barrier energy is zero. What is gained is that we do not have to deal with the large number of constraints explicitly and are hence able to solve smaller optimisation problems in every time-step where contact occurs. 
	\item Obtaining a direction of the contact force and torque that takes more information into account: by using the barrier energy, a large number of  pairs of contact points are weighted by their distance to give a net force and torque (more on this in Section \ref{measure}). 
	\end{enumerate}
An alternative formulation for the same set of constraints is $\min_i d_i\geq \hat{d}$. For a large class of optimisation methods, the interior point methods, such a problem requires a feasible starting guess, $\vec\lambda\geq\vec 0$ such that all non-overlap constraints hold, which might be very hard to obtain even for systems containing a few particles. Rewriting the non-overlap constraints in \eqref{overlap} as a barrier energy is hence necessary for robustness.
\end{remark}
\subsection{Defining the contact distance and contact force}\label{measure}
\begin{figure}[h!]
	\begin{subfigure}[t]{0.3\textwidth}
		\centering
		\includegraphics[trim = {4.7cm 0.1cm 4cm 0cm},clip,width=0.72\textwidth]{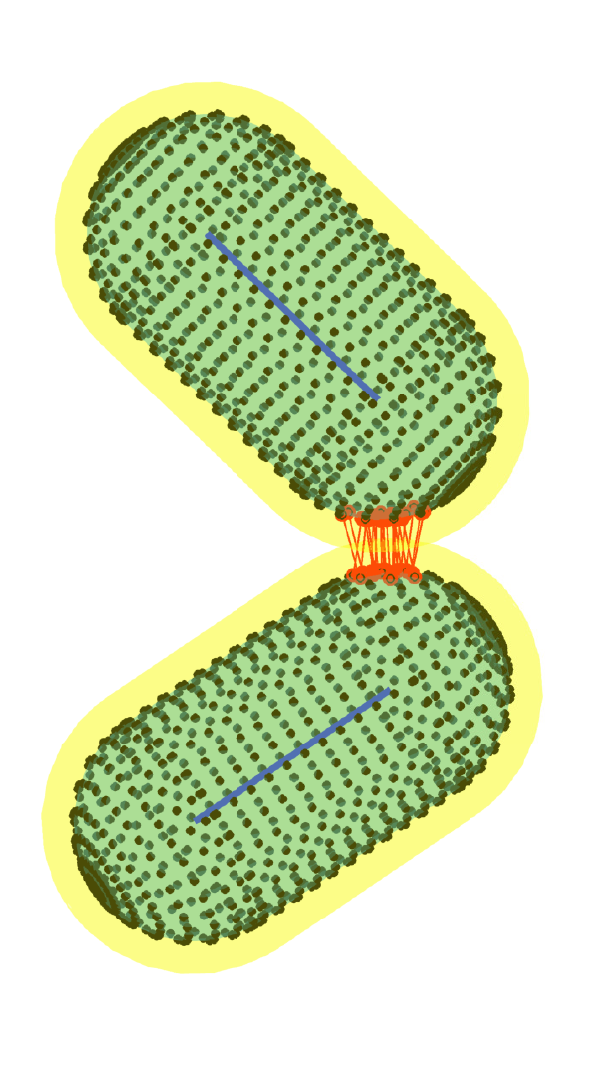}
		\caption{Contact distances $d_{ik}^{\text{s}}$ are the distances from each surface grid node on one particle (black dots) to the closest point on the other particle, less than a set threshold $\hat{d}_{\text{try}}$ (buffer region indicated in yellow). Closest points are determined from the shortest distance to the center line segment of the other particle (blue line). Pairs of grid nodes and computed points are marked with red dots, with red lines drawn in between displaying the distance.}
		\end{subfigure}~~
	\begin{subfigure}[t]{0.3\textwidth}
	\centering
		\includegraphics[trim = {4.7cm 0.1cm 4cm 0cm},clip,width=0.74\textwidth]{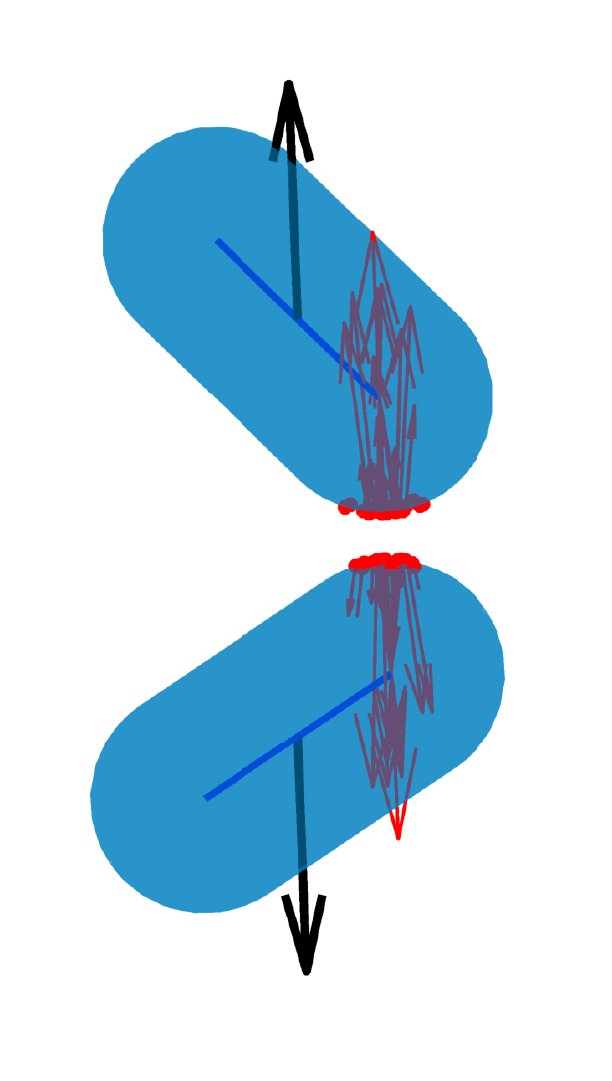}
	\caption{The corresponding directions of the contact force, representing the terms in \eqref{Ddef_C},  are visualised with red arrows scaled correlated with their contribution to the total force. There is one red arrow for each red distance line $d_{ik}^{\text{s}}$ marked in (a). The resulting contact force direction on each particle in \eqref{Ddef_C} is indicated with black thick arrows.}
\end{subfigure}~~
\begin{subfigure}[t]{0.3\textwidth}
	\centering
	\includegraphics[trim = {4.7cm 0.1cm 4cm 0cm},clip,width=0.74\textwidth]{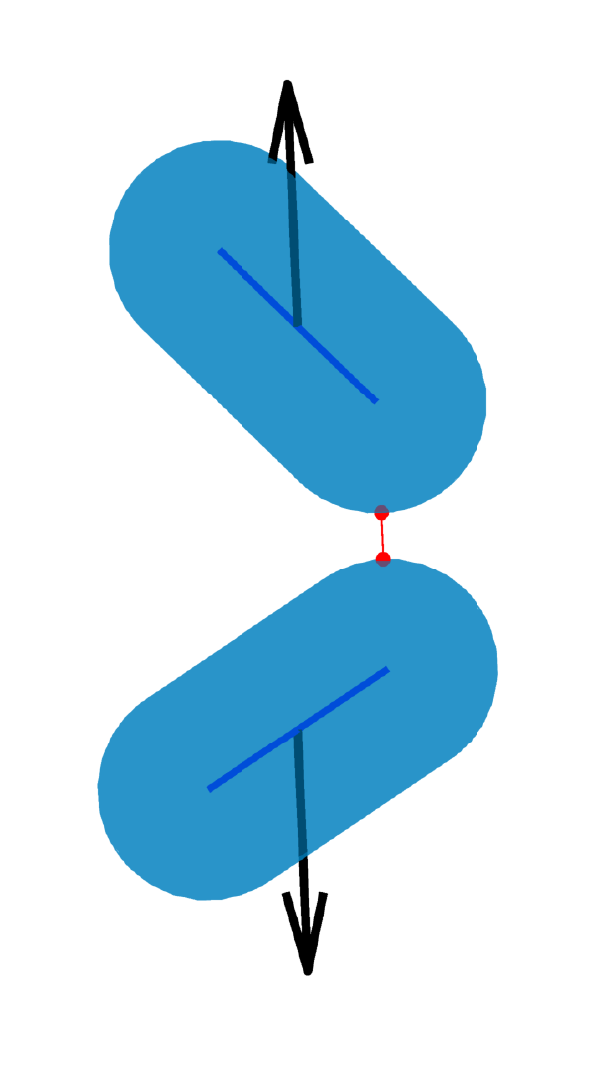}
	\caption{The shortest particle-particle distance, $d_i^{\text{p}}$, is given by the length of the red line drawn between the closest points on the two surfaces, computed from the closest distance between the two center line segments, and subtracting $2R_{\text{rod}}$. The resulting contact force direction on each particle is indicated with black thick arrows and have the same direction as the normal drawn between the contact points.}
	\label{dp_fig}
\end{subfigure}~~
\caption{Illustration of the two contact distances $d_i^{\text{p}}$ and $d_{ik}^{\text{s}}$ and the corresponding contact force directions on the particles for two fat rods close to contact in pair $i$.}
\label{contact_measures}
\end{figure}

Let us now focus on the Euclidean contact distances $d_i$ between points on the surfaces of two particles in close proximity. We make a distinction between $d^{\text{p}}$ and $d^{\text{s}}$, with  $d^{\text{p}}$ the shortest distance between particles and $d^{\text{s}}$  the shortest distance between a pair of surface points on two particles. In principle, $d_i$ in the constrained optimisation problem \eqref{minnorm} can denote either $d_i^{\text{p}}$ or $d_i^{\text{s}}$. Practically, we compute $d^{\text{p}}$ directly using known parameterisations of the particles, while the surface-point-to-surface-point distance $d^{\text{s}}$ is based on discretisations of the particle surfaces, described by a distribution of points: For each node on the surface  of one particle in the pair sufficiently close to contact, we then use the known parameterisation of the center line of the other particle to determine the closest point on this line. The particle surfaces considered in this work are all one radius away from the particle center line and the surface-point-to-surface-point distance can hence easily be computed. Geometric considerations determine if we choose to use $d^{\text{p}}$ or $d^{\text{s}}$:

\begin{enumerate}
	\item The particle-particle distance $d^{\text{p}}$  may be a good choice if the particle shapes are simple enough, such as e.g.~for spheres or axisymmetric rods with semi-spherical caps. In the latter case, we solve an equation to determine the closest distance between two line segments, following \cite{LUMELSKY1985,Ondrej2015}, and subtract $2R_{\text{rod}}$ to determine the closest distance between particles, see Figure \ref{dp_fig} for an illustration.
	\item The surface-point-to-surface-point distance, $d^{\text{s}}$, is especially beneficial in two different cases: a.) For close to parallel surfaces, such that it is hard to define a single point of contact and b.) For non-convex particles.  We will focus most of our attention on this case.

\end{enumerate}
	 

We think of the contact forces as a (hopefully small) correction to the \emph{trial} time-step $\vecmc Q_{t+\Delta t}^*$, i.e.~the time-step without contact forces, given by 
\begin{equation}
\vecmc Q_{t+\Delta t}^*=\vecmc Q_t+\Delta t\vec\Psi\vecmc U.
\end{equation}
Let $B_i(\vecmc Q_{t+\Delta t}^*,\hat{d}_{\text{try}})$ be a modified barrier energy associated with the particle contact pair $i$ aggregated over all relevant particle-particle distances at the trial time-step, with the threshold $\hat{d}_{\text{try}}$ chosen so that  $\hat{d}_{\text{try}}>\hat{d}$ ($B_i(\vecmc Q_{t+\Delta t}^*,\hat{d}_{\text{try}})$ is the sum in \eqref{barren}, but for a single particle pair and with modified threshold). Then, we define the \emph{contact force potential}, $\sum\limits_i\lambda_i B_i(\vecmc Q_{t+\Delta t}^*,\hat{d}_{\text{try}})$, a weighted sum of the barrier energy in the trial time-step, and let the contact forces be given by the gradient of this contact force potential, i.e.~
\begin{equation}\label{force_grad}
\vecmc F_c = -\nabla_{\vecmc Q_{t+\Delta t}^*}\left(\sum\limits_{i}\lambda_iB_i\left(\vecmc Q_{t+\Delta t}^*,\hat{d}_{\text{try}}\right)\right).
\end{equation} 
As stated in \eqref{eq5}-\eqref{eq6}, we may write the contact forces on the form $\vecmc F_c = \vecmc D\vec\lambda$. Let $\vec B$ be a vector of all the modified barrier energies for the $N_c$ particle pairs in contact. The sparse matrix $\vecmc D\in \mathbb R^{6N\times N_c}$, representing contact force and torque directions, depends on the geometry of the particles and their locations at the trial time-step. From \eqref{force_grad} we identify that $\vecmc D$ is given by
\begin{equation}\label{Ddefb}
\vecmc D = -\nabla_{\vecmc Q_{t+\Delta t}^*}\vec B(\vecmc Q_{t+\Delta t}^*,\hat{d}_{\text{try}}).
\end{equation}
The matrix has the structure
\begin{equation}
\vecmc D = \begin{bmatrix} \vec D_1 & \vec D_2 & \dots &\vec D_{N_c}
\end{bmatrix},
\end{equation}
where $\vec D_i$ represents contact $i$, between particle $j$ and $k$. In general, $\vec D_i$ takes the form
\begin{equation}\label{Ddef}
\vec D_i = 
\begin{bmatrix}
0~ \dots ~ 0 & \dfrac{{\vec{\hat{f}}^c_i}^T}{\|\vec{\hat{f}}^c_i\|} &  \dfrac{{\vec{\hat{t}}^c_{ij}}^T}{\|\vec{\hat{f}}^c_i\|}  & 0~ \dots ~ 0 & -\dfrac{{\vec{\hat{f}}^c_i}^T}{\|\vec{\hat{f}}^c_i\|}&  \dfrac{{\vec{\hat{t}}^c_{ik}}^T}{\|\vec{\hat{f}}^c_i\|} & 0~ \dots ~ 0 
\end{bmatrix}^T,
\end{equation}
see \cite{Tasora2008a} for a detailed motivation. From \eqref{Ddef} and the relation $\vecmc F_c = \vecmc D\vec\lambda$, it is clear that the forces  applied to the two particles in pair $i$ are  equal in magnitude (with the magnitude given by $\lambda_i$), and differ only in sign, due to Newton's third law.  In the contact force potential, all particles or grid nodes closer to each other than a tolerance $\hat{d}_{\text{try}}$ are flagged from the trial configuration  that \emph{might} come into contact in the corrected time step. All these flagged points are considered when setting up $\vecmc D$ and determines the direction of contact forces and torques.  The threshold $\hat{d}_{\text{try}}$ is chosen to set up a buffer, not to miss any colliding particles and sets at the same time the dimension of the optimisation problem, i.e.~the number of colliding particle pairs $N_c$. The contact force potential and the force definition in \eqref{force_grad} allows for contact force complementarity \emph{with respect to} $\hat{d}_{\text{try}}$ at the \emph{trial} time-step, meaning that we allow for a non-zero contact force component in a certain direction only if the associated contact distance at the trial time-step is such that $d_i(\vecmc Q^*_{t+\Delta t})<\hat{d}_{\text{try}}$.
\begin{remark}	
	We could also choose to construct the contact force potential at the previous, contact free time-step, with $\vecmc Q_{t}$. Note that this would require a larger $\hat{d}_{\text{try}}$, as particles are expected to move more in a full time-step than with only the correction from contact forces.
\end{remark}
\begin{remark}
Even if we here write the forces and torques as the gradient of a barrier energy, it  is not the gradient of a conservative potential as also $\vec\lambda$ depends on $\vecmc Q_{t+\Delta t}^*$, implicitly. 
\end{remark}

Next, we will compare two different strategies of determining the contact force potential and hence assembling the matrix $\vecmc D$, using only the very closest two points on two distinct particles or all pairs of grid nodes sufficiently close to each other. These strategies correspond to using $d^{\text{p}}(\vecmc Q^*_{t+\Delta t})$ or $d^{\text{s}}(\vecmc Q^*_{t+\Delta t})$ in the contact force potential. The differences are also outlined in Figure \ref{contact_measures}.


In the case of a contact force direction determined for the \emph{particle pair} in contact $i$, corresponding to using $d_i^{\text{p}}$, we identify that $B_i(\vecmc Q_{t+\Delta t}^*,\hat{d}_{\text{try}}) = b(d_i^p(\vecmc Q_{t+\Delta t}^*),\hat{d}_{\text{try}})$. Let $\vec n_i$ be the outward unit normal from the particle in the contact pair with the lowest index (pointing away from the contact) in the trial time-step. We let 
\begin{equation}
\begin{aligned}
\vec{\hat{f}}^c_i& =-\frac{\partial b(x,\hat{d}_{\text{try}})}{\partial d_i}\Big|_{d_i^{\text{p}}(\vecmc Q_{t+\Delta t}^*)}\vec n_i,\\ \vec{\hat{t}}^c_{ij} &= -\frac{\partial b(x,\hat{d}_{\text{try}})}{\partial d_i}\Big|_{d_i^{\text{p}}(\vecmc Q_{t+\Delta t}^*)}\left( \vec{n}_i\times \vec s_{ij}\right)\\ {\vec{\hat{t}}^c_{ik}} &= \frac{\partial b(x,\hat{d}_{\text{try}})}{\partial d_i}\Big|_{d_i^{\text{p}}(\vecmc Q_{t+\Delta t}^*)}( \vec{n}_i\times \vec s_{ik}), 
\end{aligned}
\end{equation} 
where $\vec s_{ij}$ and $\vec s_{ik}$ are the vectors from the center of each particle to the point of contact, in the global reference frame. 

We can also choose to build $\vecmc D$ for all \emph{pairs of surface points} in contact, with  $d^{\text{s}}$ the definition of separation used in the contact force potential. Let $l_j$ be the index of a grid node on particle $j$, $l_k$ an index of a grid node on particle $k$ and $\vec r_l$ denote a grid node. Then, let $\mathcal C_i$ denote the set of grid node pairs on different particles that are within a distance $\hat{d}_{\text{try}}$ from each other. Here,
the modified barrier energy $B_i(\vecmc Q_{t+\Delta t}^*,\hat{d}_{\text{try}})$ denotes the barrier energy for all grid nodes involved in the collision for the contact pair $i$, 
\begin{equation}
B_i(\vecmc Q_{t+\Delta t}^*) = \sum_{l\in\mathcal C_i} b\left(d^{\text{s}}_{l}(\vecmc Q_{t+\Delta t}^*),\hat{d}_{\text{try}}\right).
\end{equation}
Following \eqref{Ddefb}, this means that the direction of the force will be given by the sum of the normal directions for each pair of grid nodes close to contact, weighted by the corresponding derivative of the barrier function. All in all, the components of $\vec D_i$ take the form
\begin{equation}\label{Ddef_C}
\begin{aligned}
\vec{\hat{f}}^c_i &= \sum\limits_{(l_j,l_k)\in \mathcal C_i}\left(\frac{\partial b(x,\hat{d}_{\text{try}})}{\partial x}\Big|_{x = \|\vec r_{l_j}-\vec r_{l_k}\|}\right)\frac{\vec r_{l_k}-\vec r_{l_j}}{\|\vec r_{l_j}-\vec r_{l_k}\|},\\
 \vec{\hat{t}}^c_{ij} &=  \sum\limits_{(l_j,l_k)\in \mathcal C_i}\left(\frac{\partial b(x,\hat{d}_{\text{try}})}{\partial x}\Big|_{x = \|\vec r_{l_j}-\vec r_{l_k}\|}\right)\left( \frac{\left(\vec r_{l_k}-\vec r_{l_j}\right)}{\|\vec r_{l_j}-\vec r_{l_k}\|}\times\left(\vec r_{l_j}-\vec x_j\right)\right),\\
 \vec{\hat{t}}^c_{ik} &=  \sum\limits_{(l_j,l_k)\in \mathcal C_i}\left(\frac{\partial b(x,\hat{d}_{\text{try}})}{\partial x}\Big|_{x = \|\vec r_{l_j}-\vec r_{l_k}\|}\right)\left( \frac{\left(\vec r_{l_j}-\vec r_{l_k}\right)}{\|\vec r_{l_j}-\vec r_{l_k}\|}\times \left(\vec r_{l_k}-\vec x_k\right)\right).
\end{aligned}
\end{equation}
meaning that a certain force direction $\left(\vec r_{l_j}-\vec r_{l_k}\right)/\|\vec r_{l_j}-\vec r_{l_k}\|$ has a larger weight if the corresponding distance $\|\vec r_{l_j}-\vec r_{l_k}\|$ is smaller and the overlap relative to the minimum allowed distance $\hat{d}$ larger. Distances very close to $\hat{d}$ give only very small contributions to the sums representing force and torque directions in \eqref{Ddef_C}.

	 Note that alternatives could be considered for the barrier function, defining it by 
	\begin{equation}\label{b_alt}
	\quad b_0(d,\hat{d}) =  \left\{\begin{aligned}
	-\ln \left(d/\hat{d}\right),\quad &0<d<\hat{d},\\
	0,\quad &d\geq \hat{d},
	\end{aligned}\right. \quad\text{or}\quad
	b_1(d,\hat{d}) =  \left\{\begin{aligned}
	(d-\hat{d})\ln \left(d/\hat{d}\right),\quad &0<d<\hat{d},\\
	0,\quad &d\geq \hat{d},
	\end{aligned}\right. \\
	\end{equation}
	with regularity $C_0$ and $C_1$ respectively and discussed also in \cite{Li2020}. We choose the barrier function $b$ in \eqref{b_opt} due to its smooth transition at $d = \hat{d}$ ($b$ has regularity  $C_2$) which makes $b$ well-suited for the optimisation problem in \eqref{minnorm}, especially for computing derivatives of the barrier function which has to be done in any iterative method to solve \eqref{minnorm}. For determining the contact force direction at the trial time-step, we have no such regularity demand and choose the $C^0$ function $b^0$, for which the weight in \eqref{Ddef_C} from the derivative of the barrier function is the reciprocal of the distance.
	
	\begin{remark}
	Note that particles might not only violate the threshold $\hat{d}$ but also overlap in the iterative optimisation procedure before a contact force of the right magnitude is applied, such that $d_i<0$. We cannot evaluate the barrier function $b$ in \eqref{b_opt} with a negative argument. One solution is to map $d_i\to d_i+\epsilon_{\text{reg}}$ and $\hat{d}\to\hat{d}+\epsilon_{\text{reg}}$, with $\epsilon_{\text{reg}}$ a parameter to be chosen. Note that if $\epsilon_{\text{reg}}$ is chosen large, the gradient of the barrier function is small also for small $d_i$ relative to $\hat{d}$, which might have an impact on the performance of solving the optimisation problem \eqref{minnorm}. Note also that in a Brownian setting, we may need to choose a large $\epsilon_{\text{reg}}$ if a large time-step size is used (depending on how repelling the conservative potential is that gives rise to external forces and torques). A different idea to avoid a large $\epsilon_{\text{reg}}$ is to map all  $d_i < \epsilon_{\text{cap}}$ linearly such that the barrier function at $d_i = \epsilon_{\text{cap}}$ is $C^1$ and extended to a linear function, with a typical choice of $\epsilon_{\text{cap}}$ being a small fraction of $\hat{d}$, e.g. $\epsilon_{\text{cap}} = 10^{-2}\hat{d}$. Such an extension is visualised in Figure \ref{cap_reg}. The three alternative  barrier functions in \eqref{b_opt} and \eqref{b_alt} and their regularisations are visualised in Figure \ref{barrier_curve}. In numerical experiments in Section \ref{sec:Results}, we employ the strategy with $\epsilon_{\text{reg}}$, as we numerically have observed a reduced number of iterations for solving the optimisation problem with this choice, as gradients of $b$ are smaller and hence easier to handle. 
	\begin{figure}[h!]
		\centering
		\begin{subfigure}[t]{0.35\textwidth}
			\centering
			\includegraphics[trim = {1.2cm 19.2cm 13.5cm 3cm},clip,width=1\textwidth]{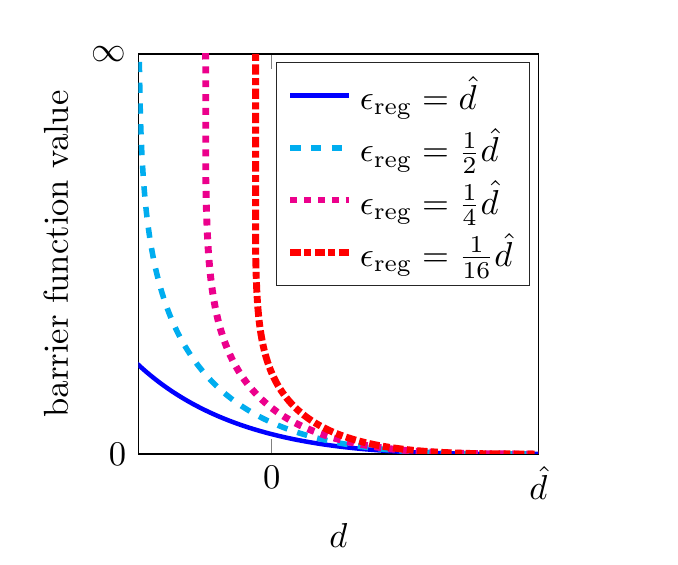}
			\caption{Modification of the barrier function $b$, where negative inter-particle distances are allowed corresponding to a maximum overlap of size $\epsilon_{\text{reg}}$.}
		\end{subfigure}~~
		\begin{subfigure}[t]{0.35\textwidth}
			\centering
			\includegraphics[trim = {1.2cm 19.4cm 13.5cm 3cm},clip,width=1\textwidth]{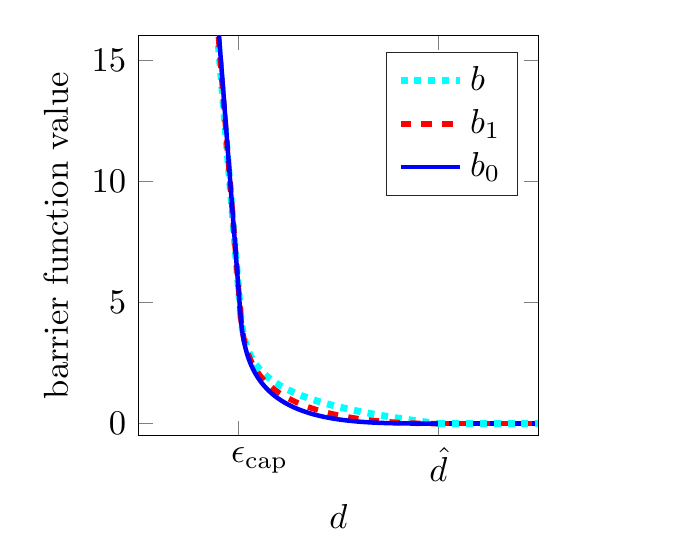}
			\caption{Alternative modification of the barrier function $b$ and its cousins $b_0$ and $b_1$, where a particle overlap with $d<\epsilon_{\text{cap}}$ is mapped to a linear continuation with $C^1$ regularity at $d = \epsilon_{\text{cap}}$.}
			\label{cap_reg}
		\end{subfigure}
		\caption{Barrier functions as defined in \eqref{b_opt} and \eqref{b_alt} are smooth approximations of the indicator function. Here, the barrier function is modified to be defined for negative inter-particle distances, as particles might overlap in the trial time-step and in iterations of the optimisation method before contact forces of the right magnitudes are found.}
		\label{barrier_curve}
	\end{figure}
\end{remark}

\subsection{Connection to a complementarity formulation}\label{comp_problem}
For comparison to complementarity techniques in the literature, as in \cite{Lu2019,Yan2019,Yan2020,Bystricky2020,Yan2022,Kohl2021}, the accumulated non-overlap constraints in $\sum\limits_i B_i(\vec\lambda) = 0$ can equivalently be expressed as a complementarity problem for the barrier energy of each contact pair and the corresponding contact force magnitude:
\begin{equation}\label{barrier_comp}
\vec 0\leq-\vec B(\vecmc Q^*_{t+\Delta t}+\Delta t\vec\Psi\vecmc M\vecmc D\vec\lambda,\hat{d}) \perp \vec\lambda \geq \vec 0.
\end{equation}
(despite the barrier function $b$ never taking negative values). The problem in \eqref{barrier_comp} can be solved by first linearising about the trial time-step so that
\begin{equation}\label{LCP}
\vec 0\leq-\vec B(\vecmc Q^*_{t+\Delta t},\hat{d})+\left(\nabla^T_{\vecmc Q_{t+\Delta t}^*}\vec B\left(\vecmc Q_{t+\Delta t}^*,\hat{d}\right)\Delta t\vec\Psi\vecmc M\nabla_{\vecmc Q_{t+\Delta t}^*}\vec B\left(\vecmc Q_{t+\Delta t}^*,\hat{d}_{\text{try}}\right)\right)\vec\lambda\perp\vec\lambda\geq\vec 0
\end{equation}
and techniques for LCPs may be used as in \cite{Fischer1992}. To guarantee a solution to the original nonlinear problem in \eqref{barrier_comp}, a sequence of linear problems of the form in \eqref{LCP} have to be solved, each with an update of the trial time-step, $\vecmc Q^*_{t+\Delta t}$, containing already computed position updates by an accumulated contact force.  
 
The equation in \eqref{barrier_comp} is reduced to the formulation by Yan et al.~in \cite{Yan2019, Yan2020, Yan2022} if only one point is involved in the contact per particle per contact pair so that
\begin{equation}
\vec 0\leq\vec d^p(\vecmc Q_{t+\Delta t}^*(\vec\lambda))-\hat{d} \perp \vec\lambda \geq\vec 0.
\end{equation}
The only difference is then that we view contact forces as a correction to a trial time-step at $t+\Delta t$ and linearise about $\vecmc Q_{t+\Delta t}^*$ instead of linearising about the previous coordinate vector $\vecmc Q_{t}$, which is done in \cite{Yan2019, Yan2020, Yan2022}. At $\vecmc Q_t$, the configuration is contact free. That is however not guaranteed at $\vecmc Q_{t+\Delta t}^*$, which likely affect the speed of convergence of the optimisation problem, as we are likely to start with an infeasible initial guess. In this work, in contrast to the work by Yan et al, we find a solution to the nonlinear complementarity problem in \eqref{barrier_comp} by computing a solution vector $\vec\lambda$ satisfying $\sum\limits_i B_i(\vec\lambda) = 0$, instead of considering a linearised problem. Note that it is not possible to strictly solve \eqref{barrier_comp}, as it is possible to have a configuration with $\vec B(\vecmc Q^*_{t+\Delta t}+\Delta t\vec\Psi\vecmc M\vecmc D\vec\lambda,\hat{d}) = \vec 0$ and $\vec\lambda = \vec 0$ and hence we allow for some relaxation in the complementarity condition. We leave it to later work for a more thorough comparison between the nonlinear barrier method presented in this paper and versions of solving linear complementarity problems if the form in \eqref{LCP} and as discussed in \cite{Lu2019,Yan2019,Yan2020,Bystricky2020,Yan2022,Kohl2021}.

\begin{remark} 
	With the choice $\hat{d}_{\text{try}} = \hat{d}$, methods for Quadratic Programming problems (QPs) are eligible to solve \eqref{LCP} (the matrix in the LCP in \eqref{LCP} is symmetric positive semi-definite \cite{Yan2019}) such as in \cite{Dai2005,Fletcher2005}. However, with such a small choice of the buffer region at the trial time-step (implied by choosing $\hat{d}_{\text{try}}$ to be as small as the natural choice of $\hat{d}$), there is a risk of missing potential contacts in the corrected time-step, leading to a long sequence of linear complementarity problems to be solved to eventually converge to a solution to the nonlinear complementarity problem. On the other hand, if $\hat{d}$ is instead increased to be as large as the natural choice of $\hat{d}_{\text{try}}$, to get equality between the two parameters, we would have a very large buffer region around each particle at the corrected time-step, with resulting contact forces heavily affecting the physics of the suspension.
\end{remark}

If the dissipative energy induced by contact forces is chosen as objective function, the Lagrangian in our nonlinar optimisation problem \eqref{minnorm} takes on a similar form as in the works by Yan et al., in which a QP on the form 
  \begin{equation}
 \min\limits_{\vec\lambda\geq \vec 0}\quad\vec\lambda^T\vecmc D^T\vecmc M \vecmc D\vec\lambda+\vec G^T\vec\lambda,
 \end{equation}
with $\vec G$ a specified vector is considered. For this QP, the first order optimality conditions is an LCP for the contact distances at the known time-step $\vecmc Q_t$ and the force magnitudes $\vec\lambda$. The difference in this paper is that we do not consider a linearised complementarity problem and the second term in the Lagrangian of \eqref{minnorm} is nonlinear and corresponds to the nonlinear constraints. Solving the problem becomes harder, but in contrast to the work by Yan et al., non-overlaps can be guaranteed.
\subsection{Solution strategy}\label{sol_strat}
The problem \eqref{minnorm} is solved with \texttt{fmincon}, an inbuilt solver for constrained minimisation in \textsc{Matlab}, employing an interior-point method. In this and any other iterative method that could be considered for solving the problem, a full Stokes solve is required for each evaluation of the barrier energy, requiring as much work as one time-step without contact handling. A large number of such iterations hence become intractably expensive and it is important to keep the number of iterations as low as possible. This number is determined by a set of parameters that have to be carefully set. The time-step size, $\Delta t$, is set a priori depending on the typical magnitudes of the computed rigid body velocities in the problem, which depend on the type of background flow and the magnitude of applied forces and torques, so that the spatial updates of the particles per time-step are reasonable in magnitude. To demonstrate the robustness of the method, we will vary the threshold  $\hat{d}$, determining the minimum allowed distance between particles and the non-zero contribution to the barrier energy at the new time-step. In a general application, we recommend to pick $\hat{d}$ as a fraction of the particle radius, e.g.~$\hat{d}=10^{-2}R$. As a rule of thumb for the remaining parameters, we pick:
\begin{itemize}
\item The stopping criterion for minimising contact force magnitudes, 
\begin{equation}
\max\limits_i|\lambda_i^{k+1}-\lambda_i^k|<10^{-2}\|\vec\lambda\|_{\infty}.
\end{equation}
\item The threshold $\hat{d}_{\text{try}}$ determining \emph{both} the $N_c$ particle pairs to potentially be assigned contact forces \emph{and} the contact force and torque directions for these pairs assembled in the matrix $\vecmc D$: For rods and boomerangs given $\Delta t$, we set $\hat{d}_{\text{try}}$ adaptively in each time-step as
\begin{equation}\label{dtry}
\hat{d}_{\text{try}} = \hat{d}+\Delta t\max_{i = 1,\dots,N}\left(\vec u_i+(L/2)\vec v_i\times\vec\omega_i\right),
\end{equation}
where the expression in the parenthesis is the tip velocity of a particle, with $\vec v_i$ the unit direction along the axis of particle $i$ and $L$ the particle length. For spheres, we pick \begin{equation}\label{sphere_try}
\hat{d}_{\text{try}} = \hat{d}+\Delta t\max\limits_{i = 1,\dots,N}\vec u_i.
\end{equation}
\item The regularisation parameter $\epsilon_{\text{reg}}$ for the barrier function as $\epsilon_{\text{reg}} = \hat{d}_{\text{try}}$.
\end{itemize}

Results are reported in Section \ref{sec:Results}. The influence of these parameter choices will specifically be considered in Section \ref{hyper}.

\section{Numerical results }\label{sec:Results}
We perform numerical experiments with spheres, rods and boomerangs. It is difficult to study the performance of the contact algorithm dynamically, as a small error in the time-discretisation or small differences in the contact forces  might result in completely different particle trajectories after sufficiently long time. No standard benchmarks are available for systems of multiple particles. For this purpose, we choose to focus on two different types of tests:
\begin{enumerate}
	\item Investigating the ability of the contact resolution strategy to find contact force magnitudes that maintain the allowed distance $\hat{d}$. This is a test of the complementarity condition stating that a non-zero force should be applied if and only if the contact is ``active''. The test is performed for geometries where the particles are pushed to come into contact by an external force, starting from a contact free configuration.
	\item Investigating the ability to preserve symmetries in a background flow that naturally enhance particle interactions to quantify the impact of different choices of $\hat{d}$ and $\Delta t$.
\end{enumerate}
In some examples, we will exaggerate $\Delta t$ and/or force magnitudes and magnitudes of background flows to trigger the contact resolution algorithm with the purpose of demonstrating its robustness.

\subsection{Spheres}\label{sphere_sec}
\subsubsection{A random suspension of spheres}\label{random_spheres_sec}
\begin{figure}[h!]
	\centering
	\begin{subfigure}[t]{0.3\textwidth}
		\centering
		\includegraphics[trim = {3cm 0.4cm 2.5cm 1cm},clip,width=1\textwidth]{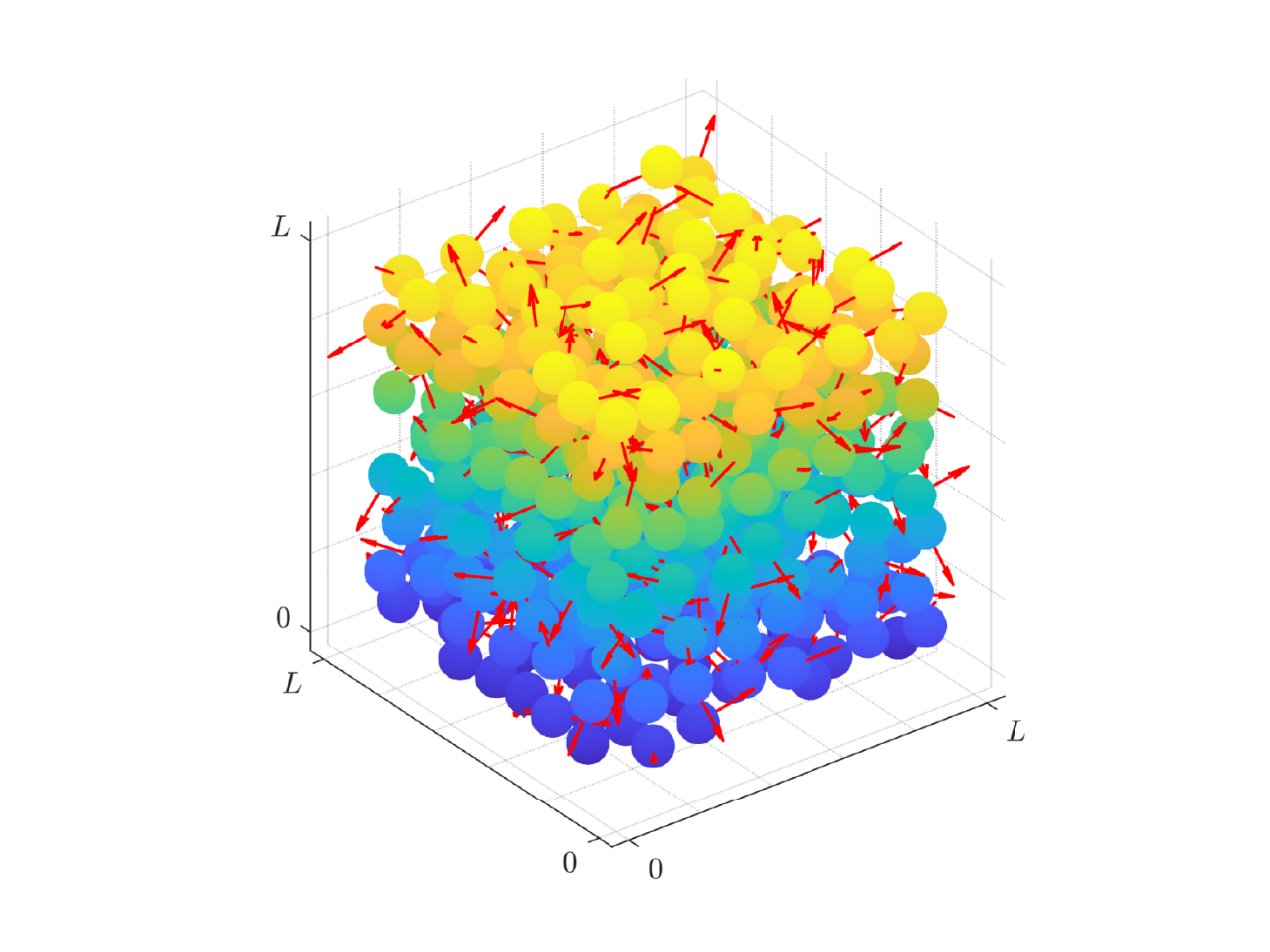}
		\caption{An example configuration with a packing density of $24\%$, with arrows indicating randomly sampled external forces pushing the particles together.}
		\label{spheres_config}
	\end{subfigure}~
\begin{subfigure}[t]{0.3\textwidth}
	\centering
	\includegraphics[trim = {3cm 0.4cm 2.5cm 1cm},clip,width=1\textwidth]{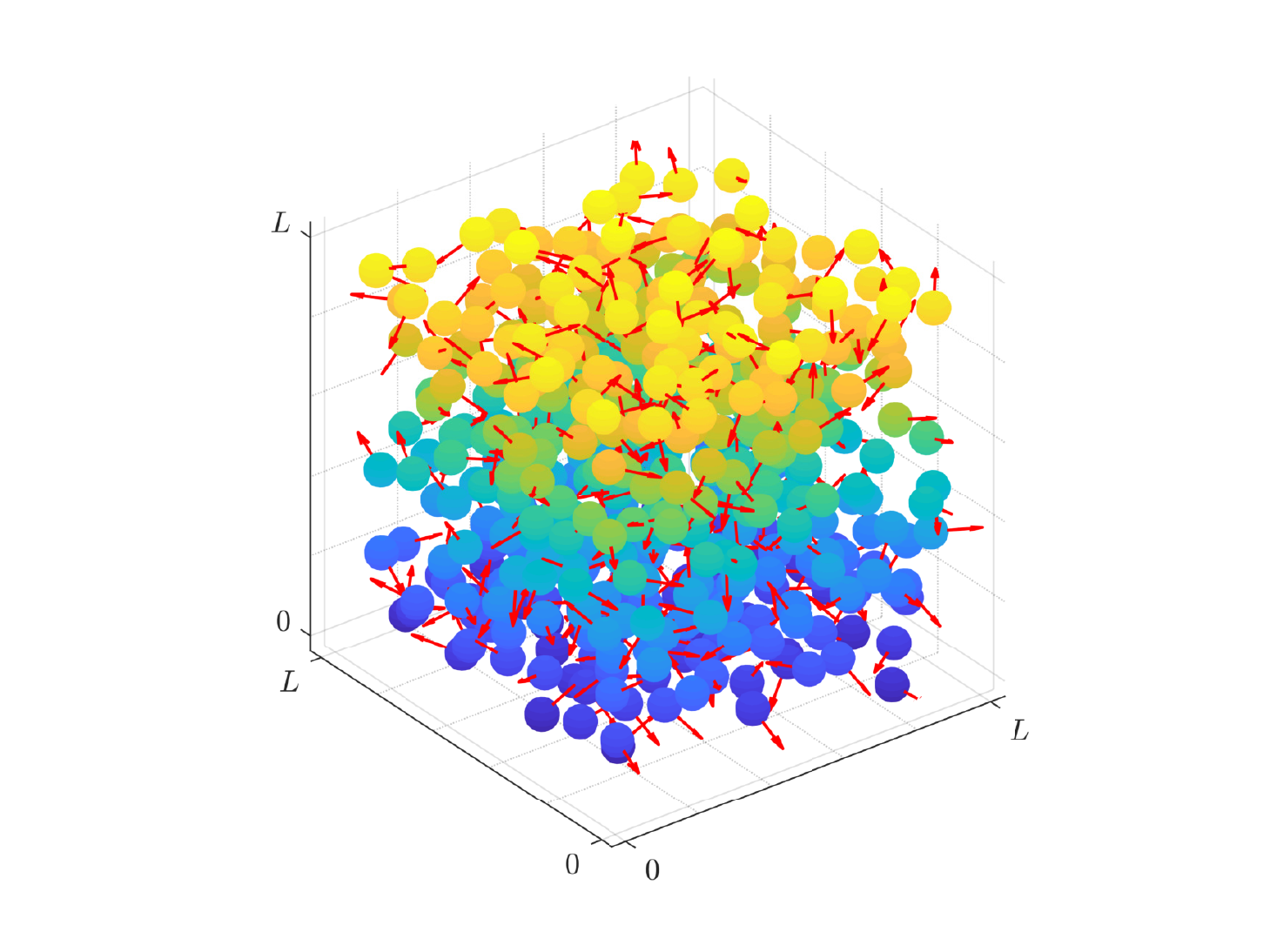}
	\caption{As in (a), but with a packing density of $12\%$.}
	\label{spheres_config2}
\end{subfigure}~
	\begin{subfigure}[t]{0.3\textwidth}
	\centering
	\includegraphics[trim = {2.5cm 18.5cm 10.0cm 2cm},clip,width=1.2\textwidth]{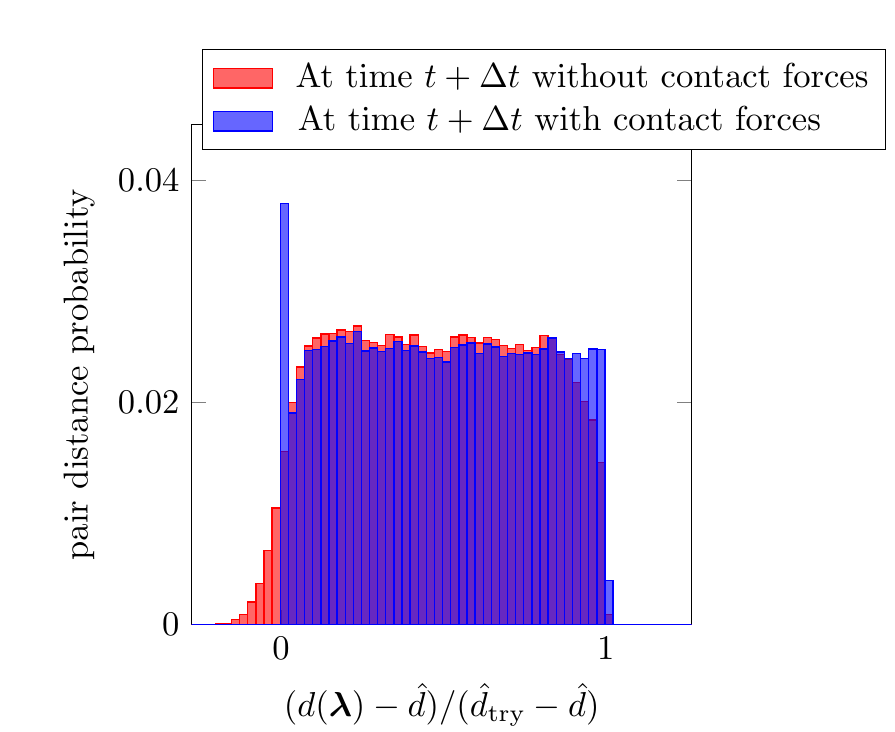}
	\caption{Inter-particle distances for the $N_c$ contact pairs before and after applying contact forces, with statistics collected from the 200 configurations with packing density $24\%$. For the size of $\hat{d}_{\text{try}}$ relative to $\hat{d}$, see panel (d).}
	\label{dist_hist}
\end{subfigure}\\
	\begin{subfigure}[t]{0.56\textwidth}
	\centering
	\includegraphics[trim = {2.5cm 19cm 6cm 2.5cm},clip,width=1.02\textwidth]{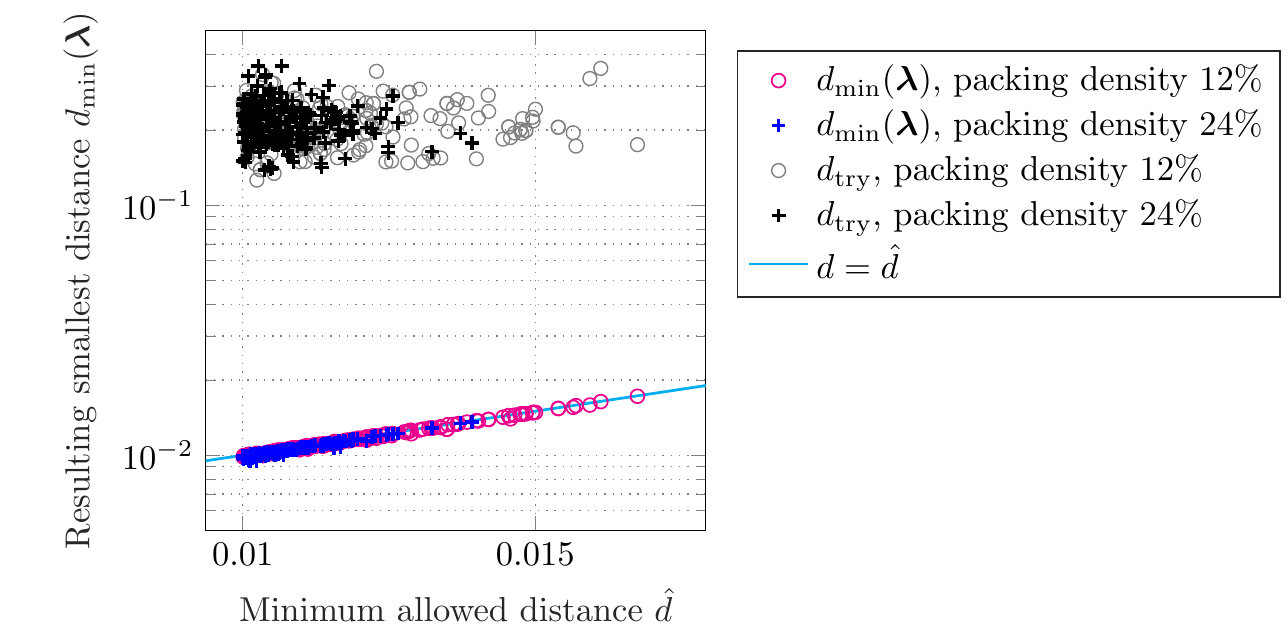}
	\caption{Minimum distances upon applying contact forces, $d_{\min}(\vec \lambda)$, respect the set distance $\hat{d}$ both for the denser and coarser configurations. For each configuration, the corresponding $\hat{d}_{\text{try}}$ is also displayed, determining the $N_c$ particle pairs flagged to be part of the contact force optimisation.}
	\label{smallest_spheres}
\end{subfigure}~
	\begin{subfigure}[t]{0.3\textwidth}
	\centering
	\includegraphics[trim = {2.5cm 18.5cm 10.0cm 2cm},clip,width=1.2\textwidth]{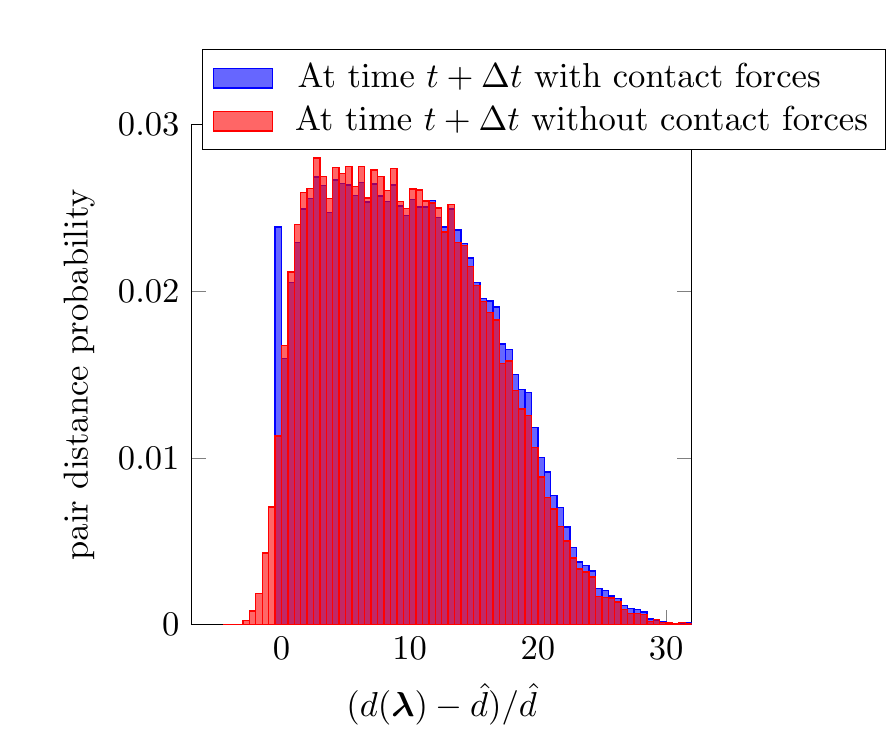}
	\caption{Same as in (c) but with distances scaled relative to $\hat{d}$ instead of $\hat{d}_{\text{try}}$.}
	\label{dist_hist2}
\end{subfigure}\\
	\caption{Contact forces are computed for all particle pairs that violate $\hat{d}$ among 500 randomly positioned spheres in a cube, where the packing density is $24\%$ and $12\%$ respectively. The spheres are affected by external forces with directions randomly drawn from the unit sphere and $\|\vec f_i\| = 100$ and a single time-step is taken with $\Delta t = 0.01$. At the trial time-step, particles are in contact if they are closer to each other than $\hat{d}$, here picked for each random configuration as the minimum separation distance at the previous, contact-free time-step. Minimum separation distances $\hat{d}$ are respected with the contact forces, and moreover, the forces do not drastically change the inter-particle distances for the closest particles, meaning that the forces are not unnecessarily large.}
	\label{random_spheres}
\end{figure}

We perform an experiment with configurations of 500 non-overlapping unit spheres randomly distributed in a cube of length $L$. The geometry is exemplified in Figures \ref{spheres_config} and \ref{spheres_config2} for the packing densities $24\%$ and $12\%$. In each of 200 configurations for each density,  $\hat{d}$ is set to be the minimum separation distance. The spheres are then assigned external forces uniformly sampled from a sphere and scaled so that $\|\vec f_i\|=100$. A trial time-step with $\Delta t = 0.01$ is taken and for spheres that have come too close, contact forces are computed with $d_i$ determined at the particle level ($d_i^{\text{p}}$ is used). Among all the corrected particle pairs, the smallest distance at the next time-level is reported vs $\hat{d}$ in Figure \ref{smallest_spheres}. In Figure \ref{dist_hist} and \ref{dist_hist2}, normalised distances are displayed for the particles flagged to potentially be in contact during the time-step, before and after applying the contact forces and scaled relative to $\hat{d}_{\text{try}}$ and $\hat{d}$ respectively. In conclusion, the forces do not artificially push the network of particles apart, but approximately keep the inter-particle distances, except for the particle-pairs that have come too close, for which the ``overlap'' is avoided by applying a contact force. 
\clearpage 
\clearpage
\subsubsection{Hyperparameter robustness test}\label{hyper}
\begin{figure}[h!]
	\centering
		\begin{subfigure}[b!]{0.65\textwidth}
		\centering
		\includegraphics[trim = {2.5cm 17.4cm 6cm 2.5cm},clip,width=0.87\textwidth]{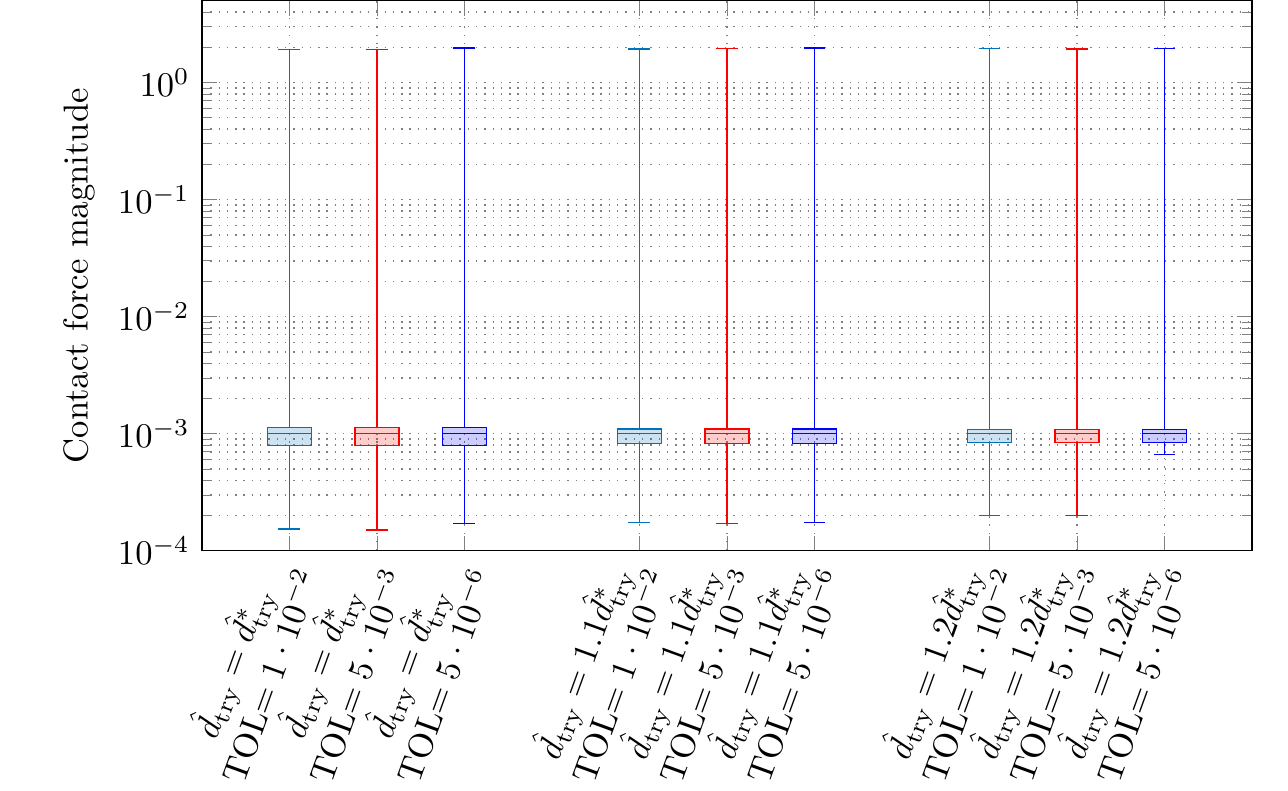}
		\caption{Bars display statistics for the contact force magnitudes with different buffer regions as given by $\hat{d}_{\text{try}}$ and stopping criteria $\max\limits_i|\lambda_i^{k+1}-\lambda_i^k|<\text{TOL}\|\vec\lambda\|_{\infty}$. Whiskers show the minimum and maximum contact force magnitudes relative to $\|\vec f_i\|$, the upper box edge the top $2.5\%$ of force magnitudes, the lower box edge the bottom $1\%$ and the line inside the box, the median.  }
		\label{hyper_bar}
	\end{subfigure}~~
	\begin{subfigure}[b!]{0.34\textwidth}
		\centering
		\vspace*{1ex}
		\includegraphics[trim = {1.5cm 17.7cm 12.5cm 2.5cm},clip,width=1.0\textwidth]{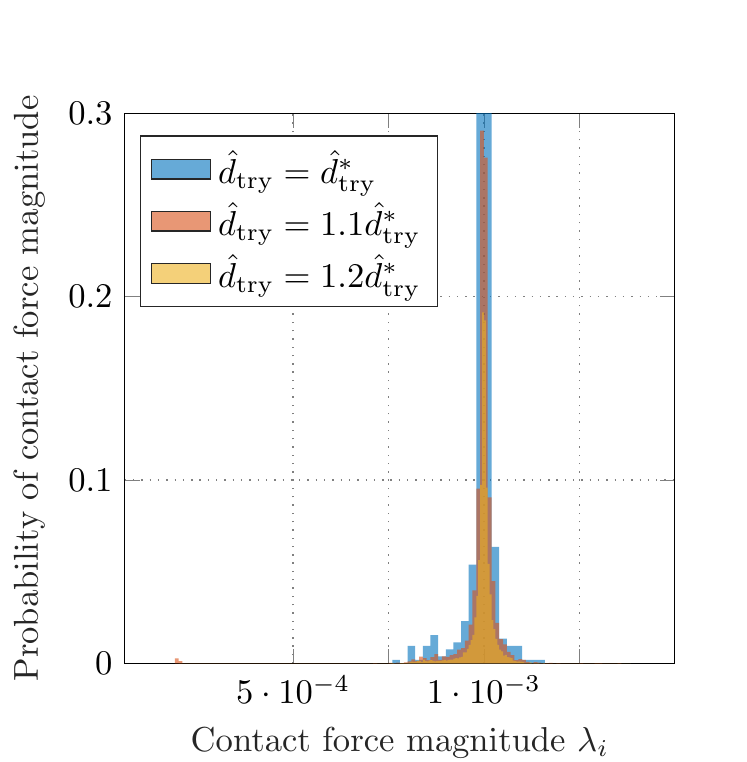}
		\caption{Histogram of the contact force magnitudes for the contact pairs with $\hat{d}_{\text{try}} > \hat{d}_{\text{try}}^*$ not flagged for collision with $\hat{d}_{\text{try}} = \hat{d}_{\text{try}}^*$. All are small in magnitude.}
		\label{hist_hyper}
	\end{subfigure}
	\caption{Almost all contact forces are small in magnitude relative to the external force, which implies that only the pair or pairs that actually violate the set minimum allowed distance $\hat{d}$ are assigned a significant contact force, even if more particle pairs are flagged to be part of the contact optimisation. For each hyperparameter combination, statistics is collected from all flagged contacts in 200 random configurations of spheres with packing density $24\%$. The same behaviour is noted for all hyperparameter combinations.}
\label{spheres_dtry}
\end{figure}

We perform the same test as for the dense suspension in Section \ref{random_spheres_sec} but vary two parameters: $\hat{d}_{\text{try}}$, that determines the buffer region around each particle in the trial time-step and the number of particle pairs flagged to potentially come in contact during the time-step, and the stopping criterion in the contact force optimisation, TOL, where the iterative optimisation method is stopped if $\max\limits_i|\lambda_i^{k+1}-\lambda_i^k|<\text{TOL}\|\vec\lambda\|_{\infty}$.  The experiment is repeated with $\hat{d}_{\text{try}}\in\lbrace \hat{d}_{\text{try}}^*,1.1\hat{d}_{\text{try}}^*,1.2\hat{d}_{\text{try}}^*\rbrace$, with $\hat{d}_{\text{try}}^*$ defined in \eqref{sphere_try} and $\text{TOL}\in\lbrace 10^{-2},5\cdot 10^{-3},5\cdot 10^{-6}\rbrace$.  By viewing the statistics in Figure \ref{hyper_bar}, one can conclude that only the elements of $\vec\lambda$ corresponding to particle pairs that would overlap without a contact force correction get assigned a larger contact force and all the other magnitudes are small, for all tested combinations of hyperparameters.  The choice of TOL will affect the magnitudes of the computed contact forces to a very small extent as long as TOL is small enough. One would expect that with a more restrictive TOL, the number of non-negligible contact forces is reduced, but on the other hand, the number of iterations required to solve \eqref{minnorm} is increased. In practice, the dependence of TOL for the magnitude of the contact foce is however very small, see Figure \ref{hyper_bar}. In Figure \ref{hist_hyper}, the histogram shows the contact force magnitudes only for the pairs flagged with $\hat{d}_{\text{try}}>\hat{d}_{\text{try}}^*$ that are not flagged with $\hat{d}_{\text{try}}=\hat{d}_{\text{try}}^*$. Even if more particle pairs are flagged within the time-step with a larger $\hat{d}_{\text{try}}$, we show that all the extra contact forces computed with a larger buffer region are small in magnitude. In Table \ref{MSE} the mean squared deviation and max relative difference of the three components of the contact force for any particle in 200 configurations are presented versus the setting with $\hat{d}_{\text{try}}=\hat{d}_{\text{try}}^*$ and $\text{TOL} = 5\cdot 10^{-6}$.

\begin{figure}[h!]
\centering
\includegraphics[trim = {1.5cm 17.7cm 12.5cm 2.5cm},clip,width=0.35\textwidth]{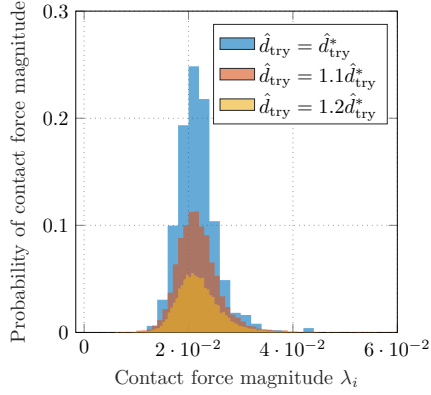}
		\caption{Histogram of the contact force magnitudes for the contact pairs with $\hat{d}_{\text{try}} > \hat{d}_{\text{try}}^*$ not flagged for collision with $\hat{d}_{\text{try}} = \hat{d}_{\text{try}}^*$ using the objective function $\vec\lambda^T\vecmc D\vecmc M\vecmc D\vec\lambda$ and the strice stopping criterion $\text{TOL}=5\cdot 10^{-6}$. A lot of these extra contact forces are not small in magnitude, which would be expected, and we therefore conclude that this choice of objective function is not as robust as $\sum_i\lambda_i$.}
		\label{extra_diss}
\end{figure}

We end this section with a discussion on the choice of objective function. If we redo the experiment with $\text{TOL}=5\cdot 10^{-6}$ and $\hat{d}_{\text{try}}\in\lbrace \hat{d}_{\text{try}}^*,1.1\hat{d}_{\text{try}}^*,1.2\hat{d}_{\text{try}}^*\rbrace$, but change the objective function to the approximated dissipative energy $\vec\lambda^T\vecmc D^T\vecmc M\vecmc D\vec\lambda$, the ``extra'' flagged particle pairs for larger $\hat{d}_{\text{try}}$ will get assigned a larger contact force magnitude, relative to choosing the 1-norm objective function. The force magnitudes for these particle pairs are reported in Figure \ref{extra_diss} and can be compared to Figure \ref{hist_hyper}.

\clearpage
\begin{table}[h!]
\centering
\begin{tabular}{c c}
\begin{tabular}{c | c |c | c}
 \multicolumn{4}{c}{(a)\textbf{ MSD of contact force }} \\ \hline
$\hat{d}_{\text{try}}$ / TOL & $1\cdot 10^{-2}$ & $5\cdot 10^{-3}$ & $5\cdot 10^{-6}$ \\ \hline
$\hat{d}_{\text{try}}^*$ & $2.45\cdot 10^{-6}$ & $6.63\cdot 10^{-7}$ & \text{Reference} \\ \hline
$1.1\hat{d}_{\text{try}}^*$ & $3.76\cdot 10^{-3}$ & $9.60\cdot 10^{-6}$& $8.84\cdot 10^{-6}$ \\ \hline
$1.2\hat{d}_{\text{try}}^*$ & $3.86\cdot 10^{-3}$& $1.03\cdot 10^{-5}$ & $9.72\cdot 10^{-6}$  \\ \hline
\end{tabular} &
\begin{tabular}{c | c |c | c}
 \multicolumn{4}{c}{(b) \textbf{Max (relative) difference of contact force }} \\ \hline
$\hat{d}_{\text{try}}$ / TOL & $1\cdot 10^{-2}$ & $5\cdot 10^{-3}$ & $5\cdot 10^{-6}$ \\ \hline
$\hat{d}_{\text{try}}^*$ & $0.14$ & $0.14$ & \text{Reference} \\ \hline
$1.1\hat{d}_{\text{try}}^*$ & $0.15$ & $0.16$& $0.16$ \\ \hline
$1.2\hat{d}_{\text{try}}^*$ & $0.16$ & $0.17$ & $0.17$  \\ \hline
\end{tabular} 
\end{tabular}
\caption{Mean squared deviation and max difference of contact forces relative to the reference with $\hat{d}_{\text{try}}=\hat{d}_{\text{try}}^*$ and $\text{TOL} = 5\cdot 10^{-6}$ with different hyperparameter combinations ($\hat{d}_{\text{try}}$, TOL) computed over all three components of the contact force in for all spheres in 200 random configurations. }
\label{MSE}
\end{table}

\subsection{Rods}
\subsubsection{Geometric considerations}\label{rod_type}
Rods constituting of a central cylindrical part and semi-spherical caps at both ends are considered. The multiblob grid is set via a matching strategy for the mobility coefficients of a single particle as described in \cite{Broms2022}. The surface of the rod is everywhere one radius from its center line segment and inter-particle distances $d^{\text{p}}_i$ can hence be determined by computing the shortest distance between two line segments and subtracting $2R_{\text{rod}}$, see \cite{LUMELSKY1985,Ondrej2015}. The surface-node-to-surface-node distances $d^{\text{s}}_{ik}$ are determined by computing the shortest distance between each source grid node belonging to an upsampled grid on particle 1 and the line segment of particle 2. A new target surface node is introduced on particle 2 where the vector of shortest distance cuts the surface of the particle. This source/target pair is added to the list of surface points that will be used to form the elements in the matrix $\vecmc D$ in \eqref{Ddef_C}. The procedure is repeated for all points sufficiently close to the other particle on the upsampled surfaces of both particles. Finally, the closest points of contact as defined by $d_i^{\text{p}}$ are added to the list not to underestimate the closest distance between the particles, as for all grid node pairs $k$, $d^{\text{s}}_{ik}\leq d_i^{\text{p}}$.
\subsubsection{A chain of rods}
\begin{figure}[h!]
	\centering	
	\begin{subfigure}[t!]{0.44\textwidth}
			\centering		
			\includegraphics[trim = {1cm 1cm 2.0cm 0.7cm},clip,width=1\textwidth]{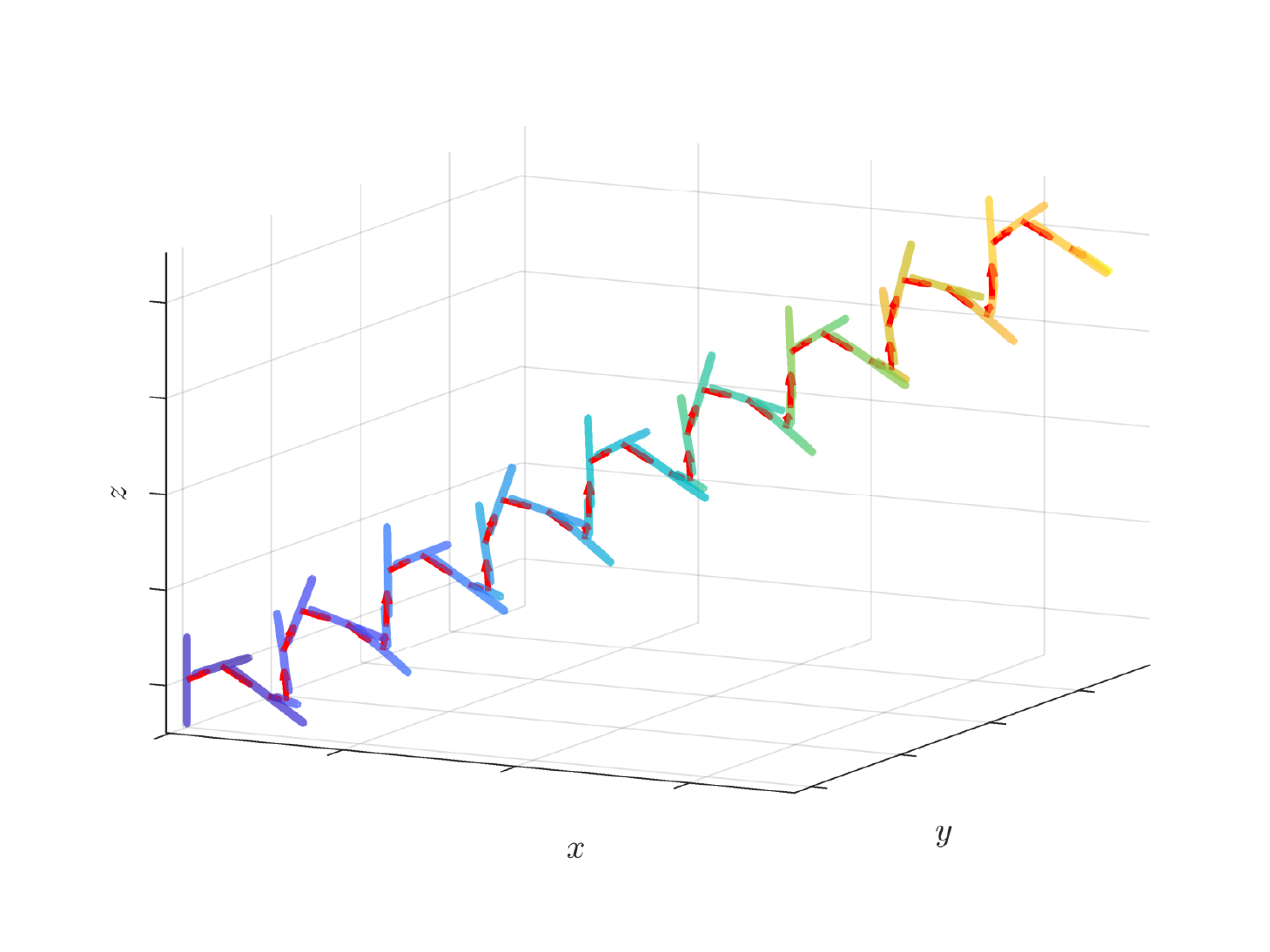}
		\caption{Example rod chains with the same relative translation and rotation between each consecutive pair of rods so that the inter-particle distance $\hat{d}$ is equal between every pair. Red arrows indicate external forces towards the center of the next particle that deliberately cause collisions during the next time-step. Particle colors indicate the depth in the suspension. }
		\label{twist_geom}
	\end{subfigure}~~
			\begin{subfigure}[t!]{0.53\textwidth}
	\centering
	\hspace*{-2ex}
	\includegraphics[trim = {2.7cm 17.2cm 6.2cm 1.5cm},clip,width=1.03\textwidth]{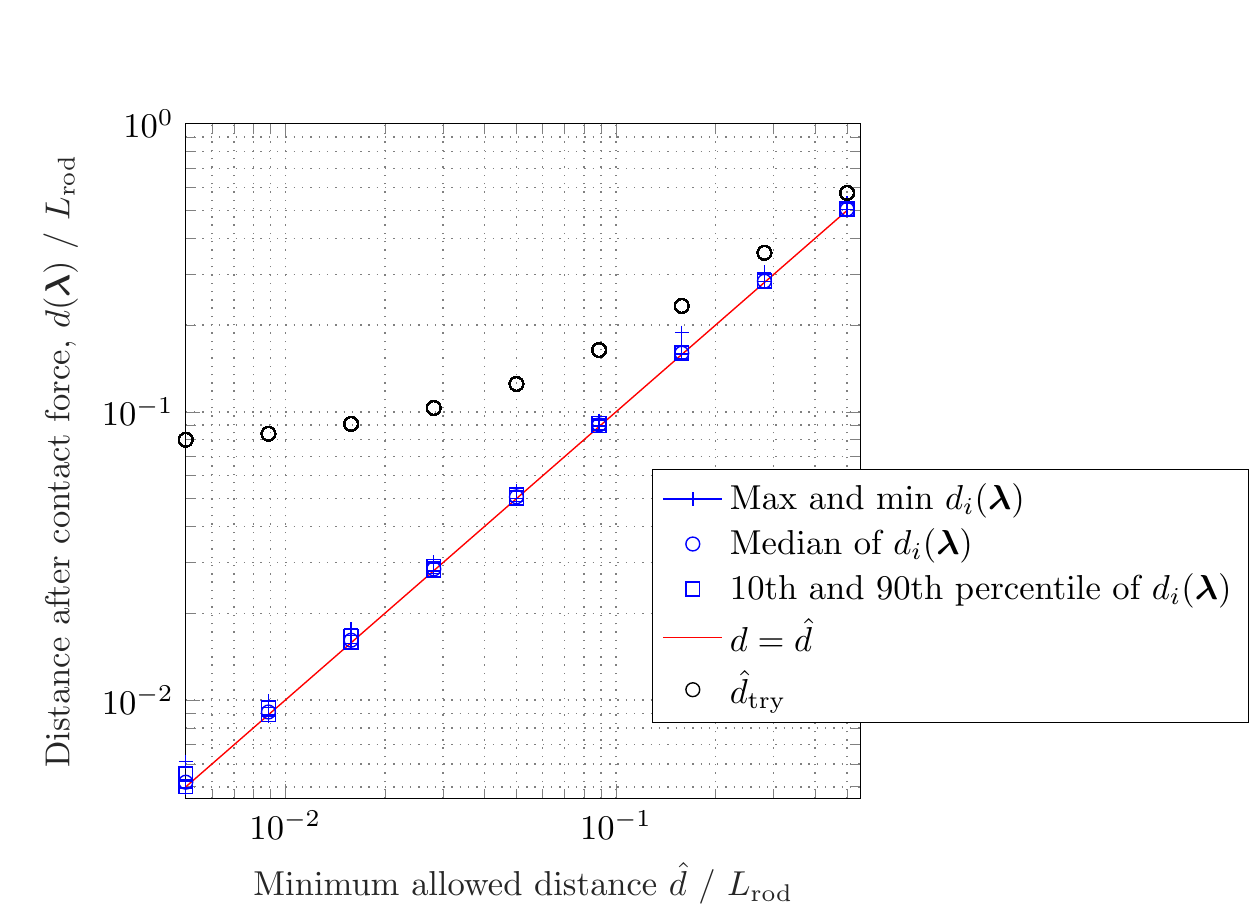}
	\caption{Distributions of the computed distance $d(\vec\lambda)$ with statistics collected from all rods in 100 chains for each $\hat{d}$, as compared to the reference minimum allowed distance $\hat{d}$. For each $\hat{d}$, the corresponding distance $\hat{d}_{\text{try}}$ is also indicated, for which particles are flagged to be part of the collision algorithm.  }
	\label{chain_error}
\end{subfigure}\\
		\begin{subfigure}[b!]{0.49\textwidth}
	\centering
	\includegraphics[trim = {2.7cm 17.3cm 8.8cm 2cm},clip,width=0.92\textwidth]{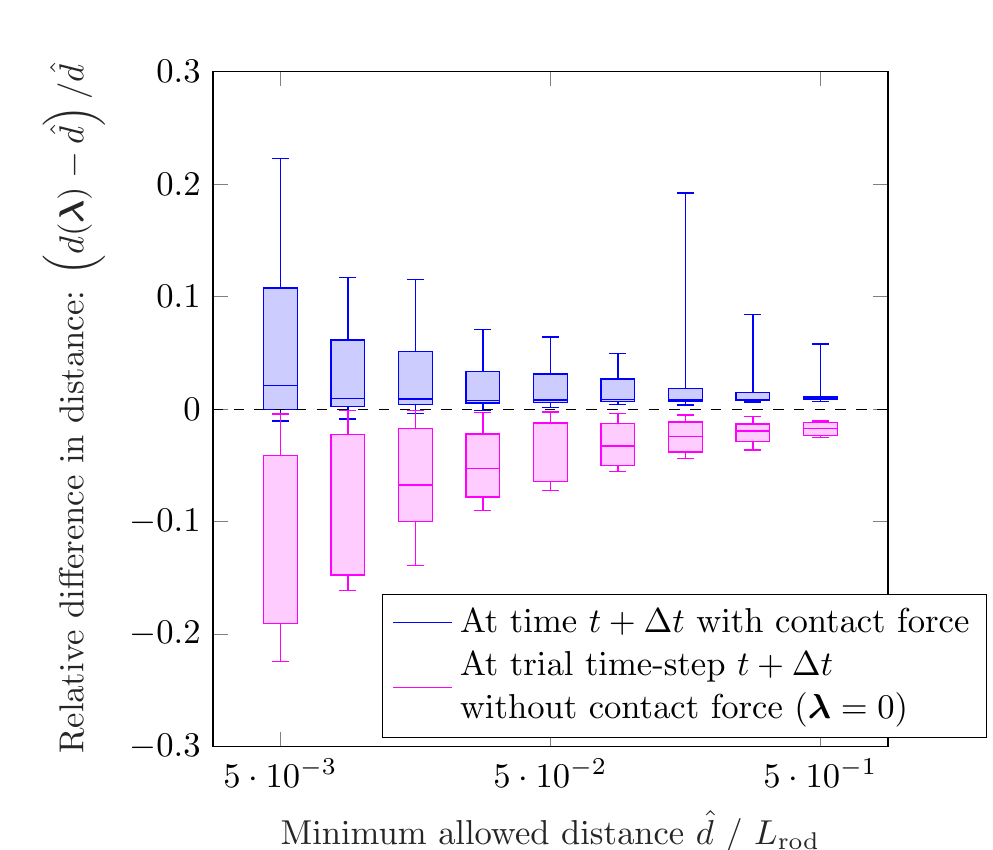}
	\caption{Bars show the relative difference in computed distances $d(\vec\lambda)$ to $\hat{d}$ at the trial time-step without contact forces and upon applying contact forces, for each choice of $\hat{d}$. Whiskers display the minimum and maximum relative distance difference, box edges the 10th and 90th percentile of the relative difference and the box center line display the median.}
	\label{chain_rel_error}
\end{subfigure}~~
		\begin{subfigure}[b!]{0.49\textwidth}
	\centering
	\vspace*{-8ex}
	\includegraphics[trim = {2.7cm 17.6cm 9.9cm 2cm},clip,width=0.8\textwidth]{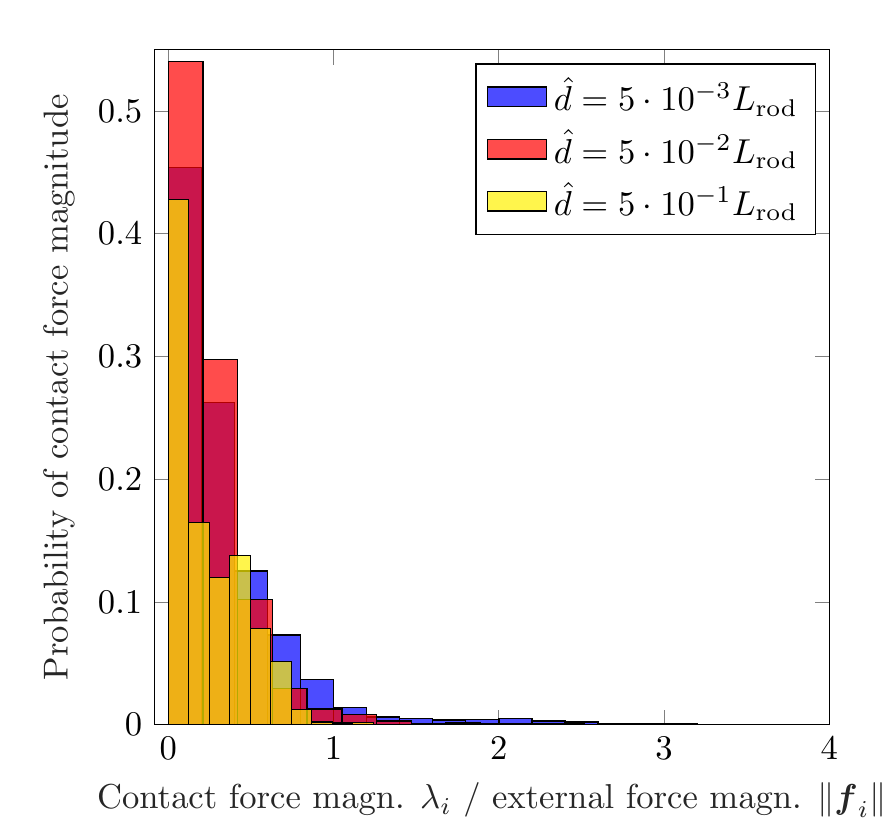}
	\caption{Magnitude of contact forces relative to the magnitude of the external forces $\|\vec{f}_i\|$. All computed contact forces are comparable in size to $\|\vec{f}_i\|$ or smaller.}
	\label{force_hist}
\end{subfigure}~

\caption{Chains of 40 rods are forced to come into contact by externally applied forces towards the next particle in the chain. By finding optimal magnitudes of the contact forces, the minimum separation distance $\hat{d}$ can be kept. For each $\hat{d}$, the experiment is repeated with 100 randomly generated chains.}
	\label{rod_chain}
\end{figure}

Chains of rods of length $L_{\text{rod}} = 0.5$ and radius $R_{\text{rod}} = L_{\text{rod}}/20$ are considered, where for each chain, a unit direction vector $\vec t\in \mathbb R^3$ and rotation $(\theta,\phi)$ are drawn at random from the first orthant. The chain is constructed with the first rod placed at the origin with orientation coinciding with the $z$-axis. The consecutive rods are obtained from the previous by rotating the particle by $(\theta,\phi)$ and then translating the center coordinate by $\beta(\hat{d})\vec t$ in the coordinate frame of the previous particle, with the constant $\beta(\hat{d})$ determining the magnitude of the translation such that the smallest distance between a pair of particles is $\hat{d}$. Example geometries are visualised in Figure \ref{twist_geom}.

Every rod in each chain is assigned a force in the direction of the next particle so that  a collision is caused in the next time-step ($\hat{d}$ is violated) if no contact forces are applied. The given set of forces on the chain configuration is then corrected with contact forces. Statistics for the resulting particle-particle distances $d(\vec\lambda)$, upon applying contact forces, are reported versus $\hat{d}$ for a large range of $\hat{d}$ in Figure \ref{chain_error}. Note that the largest $\hat{d}$  do not represent a realistic choice for a dynamic simulation, but rather demonstrate the robustness of the method and that the choices of $\hat{d}_{\text{try}}$ depend on $\Delta t$ and therefore do not follow the curve for $\hat{d}$. By comparing $d(\vec\lambda)$ to the reference distance $\hat{d}$, it can be concluded that the minimum allowed distance $\hat{d}$ is respected for almost all particles in all chains. Statistics for $\left(d(\vec\lambda)-\hat{d}\right)/\hat{d}$ is illustrated for all chains before and after contact in Figure \ref{chain_rel_error}. In Figure \ref{force_hist}, histograms display probable contact force magnitudes on any rod in any chain for fixed $\hat{d}$ relative to the magnitude of the external force triggering the contact avoiding algorithm. Contact forces are comparable in magnitude to the external forces.

For reproducibility, the time-step size is set to $\Delta t = 0.05$ and the force magnitude is $\|\vec f_i\| = 2\hat{d}^{1/3}$ (this particular choice was made to promote collisions for the entire range of $\hat{d}$ considered in the test).

\subsubsection{Rods in a biaxial compression flow}\label{biaxial}
\begin{figure}[h!]
	\begin{subfigure}[t]{0.35\textwidth}
		\centering
		\includegraphics[trim = {2cm 0cm 1cm 0cm},clip,width=1.05\textwidth]{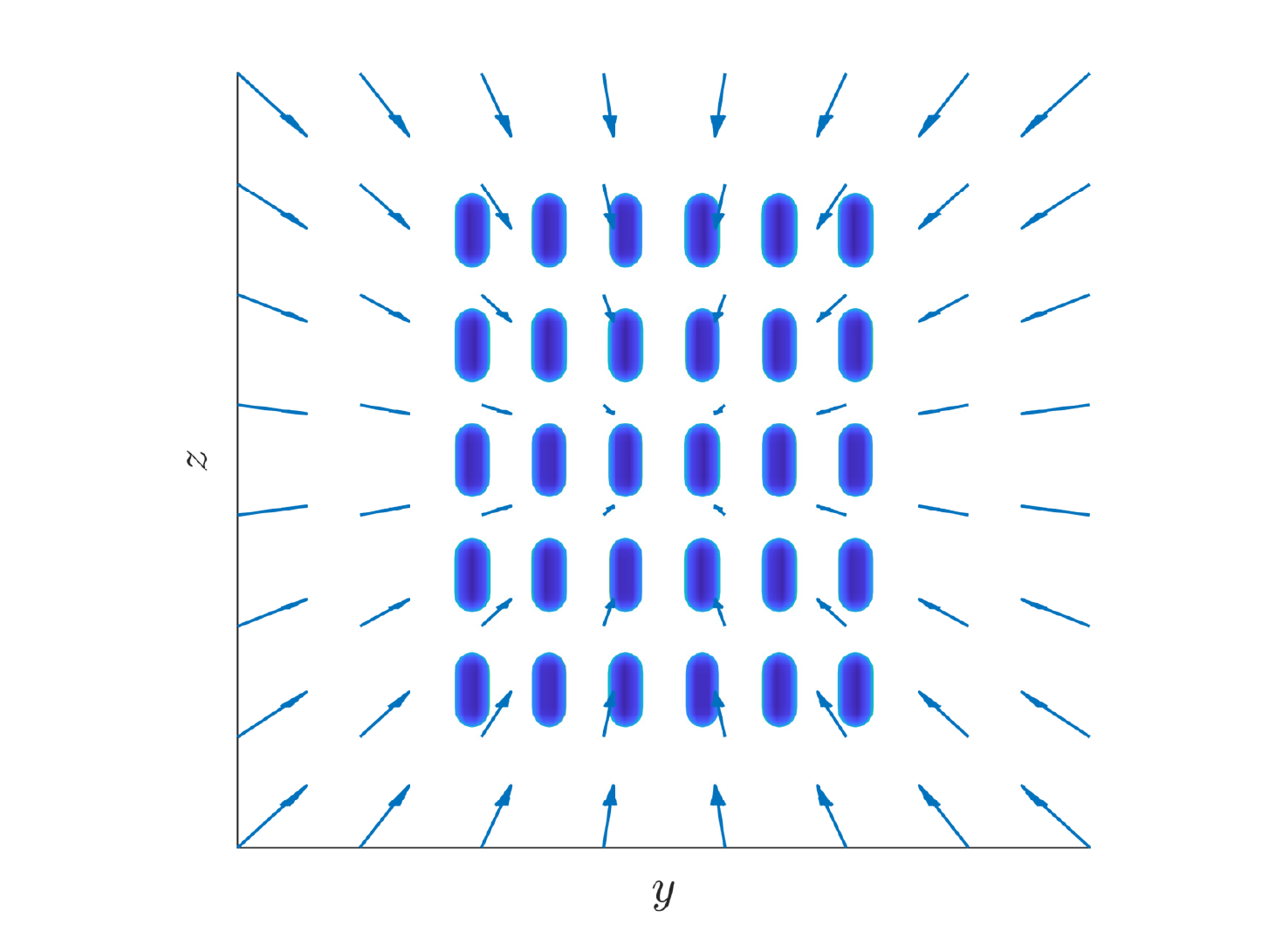}
		\caption{Initial particle geometry in the $yz$-plane. Arrows indicate flow direction.}
		\label{rod_grid1}
	\end{subfigure}~~
	\begin{subfigure}[t]{0.18\textwidth}
	\centering
	\includegraphics[trim = {4.5cm 0cm 4cm 0cm},clip,width=1.1\textwidth]{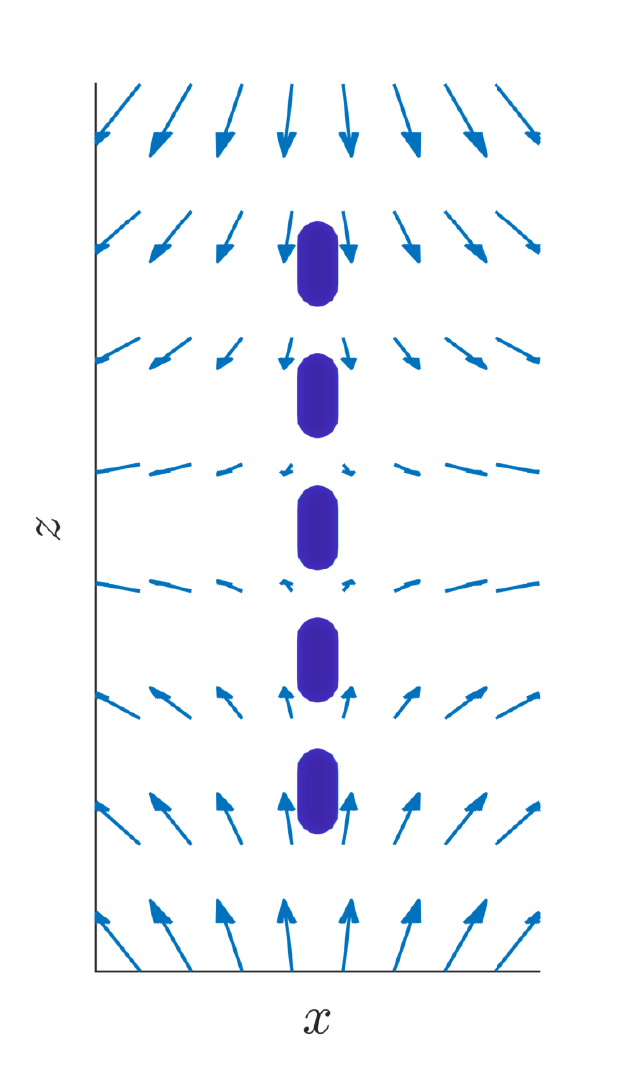}
	\caption{Initial particle geometry in the $xz$-plane. Arrows indicate flow direction.}
	\label{rod_grid2}
\end{subfigure}~~
		\begin{subfigure}[t]{0.45\textwidth}
	\centering
	\includegraphics[trim = {3cm 16.2cm 9.1cm 2cm},clip,width=0.84\textwidth]{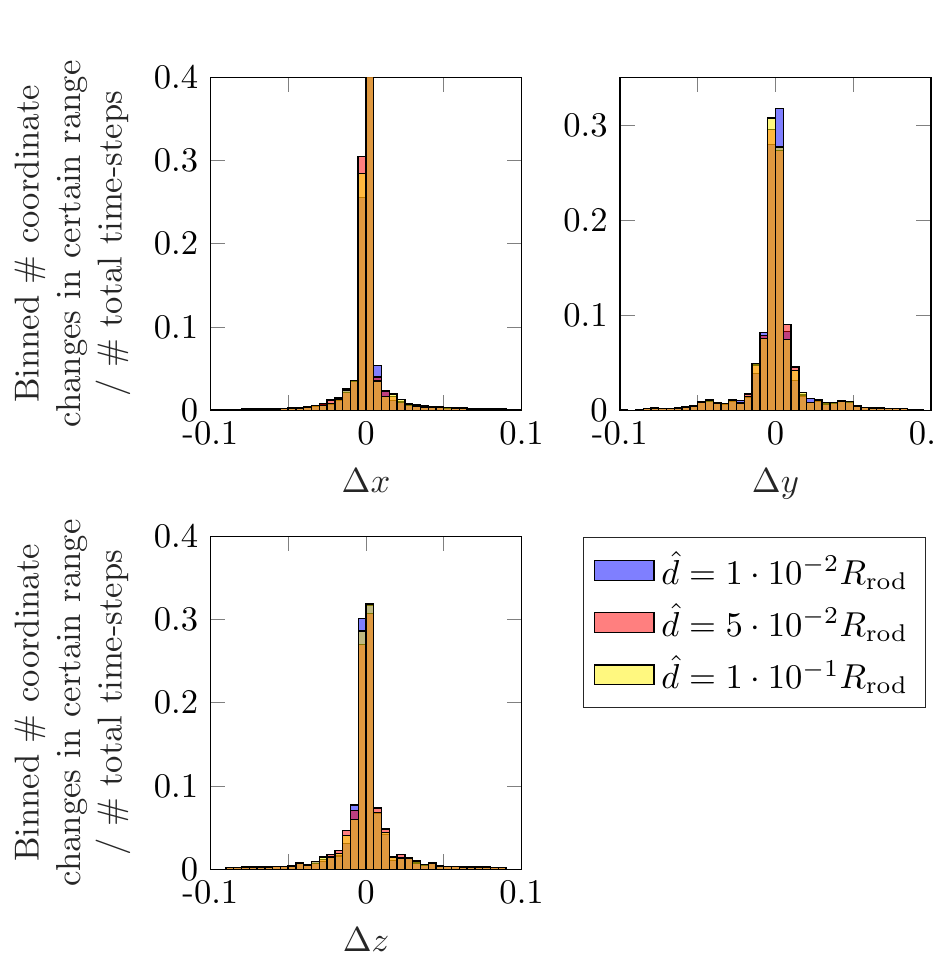}
	\caption{Binned particle coordinate change per time-step with the large time-step size $\Delta t = 0.02$. The contact forces cause no drastic jumps in the particle trajectories and all coordinate changes are small.}
	\label{biaxial_coord}
\end{subfigure}
	\begin{subfigure}[t]{0.48\textwidth}
		\centering
		\hspace*{-2ex}
				\includegraphics[trim = {3cm 18.0cm 5.0cm 2cm},clip,width=1.04\textwidth]{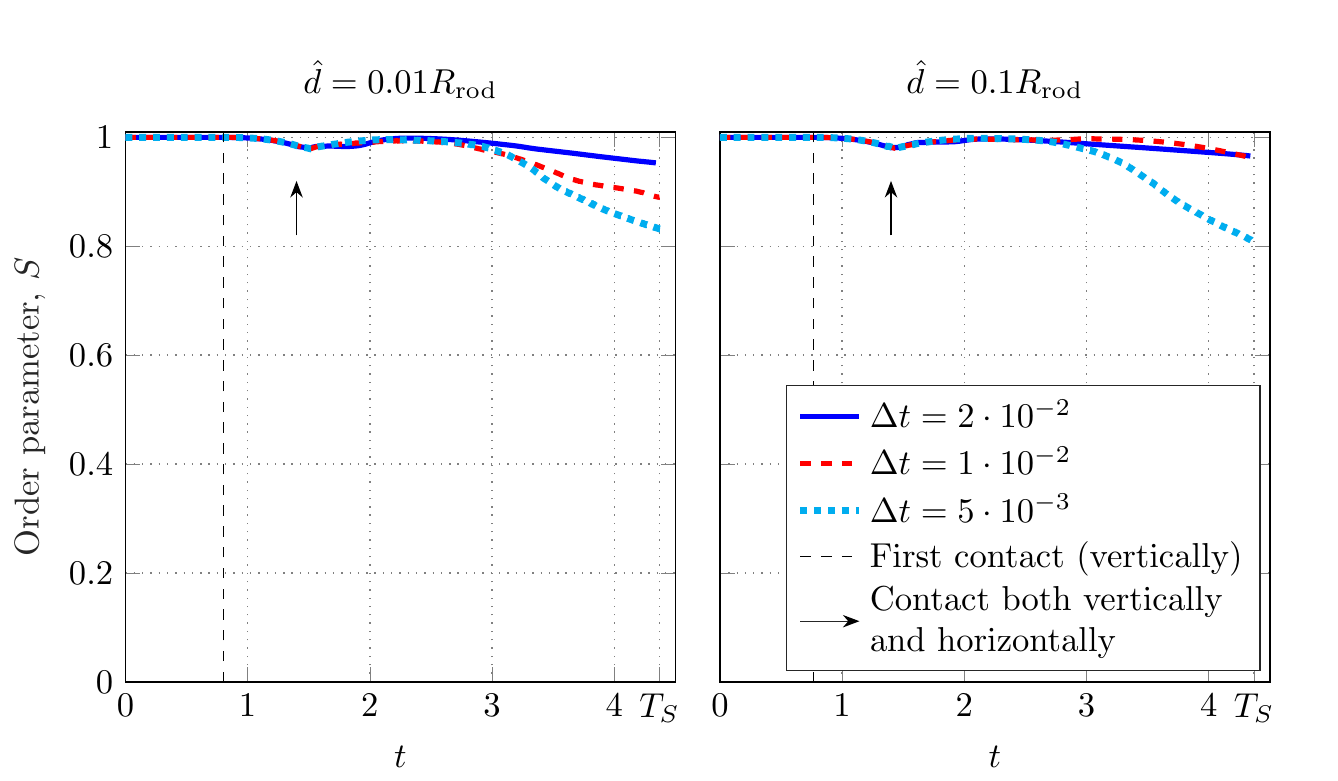}
					
			\caption{The order parameter $S$ as defined in \eqref{order_p}, using various time-step sizes and two different choices of the minimum allowed distance $\hat{d}$. When all particles are aligned with the $z$-axis, $S=1$. The alignment of the particles is broken after some finite time. With the largest $\Delta t$, this happens after approximately 100 time-steps using $\hat{d} = 0.01R_{\text{rod}}$, at $t\approx 3$.}
		\label{order_param}
	\end{subfigure}~~	
	\begin{subfigure}[t]{0.48\textwidth}
		\centering
				\includegraphics[trim = {3cm 18.0cm 5.0cm 2cm},clip,width=1.04\textwidth]{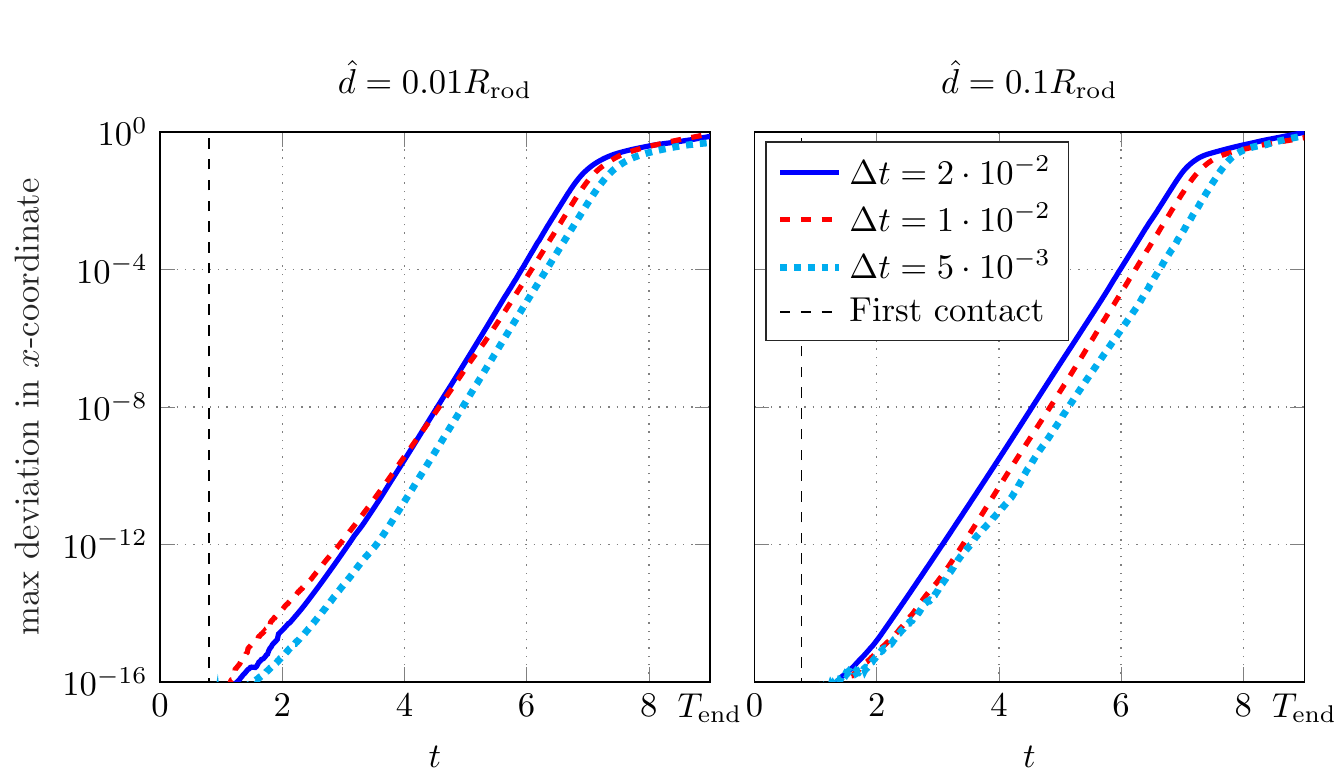}				
		\caption{Maximal deviation in the $x$-coordinate: Despite the large number of time-steps with the contact algorithm active, particles stay in the $yz$-plane for a long time. The number of time-steps taken are reported in Table \ref{tab:dt}.}
		\label{x_dev}
	\end{subfigure}
	
	\caption{A grid of 30 rods affected by a biaxial compression flow. Contact forces are robustly computed whenever the minimum allowed distance $\hat{d}$ is violated. The symmetry of the configuration is broken but this happens after a large number of time-steps. Despite the challenging setting with the background flow pushing particles together, the particles stay in the original 2D-grid for a relatively long time.}
\end{figure}

Consider a 2D grid of 30 rods of length $L_{\text{rod}}=2$ and radius $R_{\text{rod}}=L_{\text{rod}}/4$ arranged as in Figures \ref{rod_grid1}-\ref{rod_grid2} and affected by a background biaxial compression flow $\vecmc U_{\text{bg}}$ as indicated in the figure, given by
\begin{equation}
\vecmc U_{\text{bg}} = \begin{bmatrix} (\vec E\vec x_1)^T,\vec 0,(\vec E\vec x_2)^T,\vec 0,\dots,(\vec E\vec x_n)^T,\vec 0
\end{bmatrix}^T\quad\text{with }\vec E = 
	\begin{bmatrix} \dot{\gamma} & 0 & 0 \\ 0 & -\dot{\gamma}/2 & 0 \\ 0 & 0 & -\dot{\gamma}/2
	\end{bmatrix},\medspace \gamma = 1.
\end{equation} We choose different minimum allowed distances $\hat{d}$ and a hierarchy of time-step sizes $\Delta t$ and discretise the dynamics of the system with forward Euler with one hyperparameter setting $(\hat{d},\Delta t)$ at a time. Whenever the smallest allowed distance $\hat{d}$ is violated at the end of the time-step, contact forces are computed for particles considered sufficiently close to contact at the trial time-step, as given by $\hat{d}_{\text{try}}$.  Due to the contractile nature of the background flow, the contact avoiding algorithm will be triggered in almost every time-step, see Table \ref{tab:dt}. In Figure \ref{biaxial_coord}, the change in coordinate position per time-step is displayed for all particles until the particles start to diverge along the $x$-axis (where the background flow is diverging) and the simulation is stopped. The contact forces cause no drastic jumps in the particle trajectories and a vast majority of the coordinate updates are very close to zero. Table \ref{tab:dt} indicates that contact forces are robustly computed for a long sequence of time-steps. The maximum deviation along the $x$-axis for the particles is displayed as function of time in Figure \ref{x_dev}. A smaller time-step allows for a slightly smaller deviation than a larger time-step. Note however that for all hyperparameters, particles stay in the $yz$-plane.

If all symmetries of the problem were kept, the particles should not only stay in the same plane, but also keep the alignment with the $z$-axis. The alignment can be quantified with the Onsager order parameter $S$ defined by \cite{Doi2019}
\begin{equation}\label{order_p}
S = \frac{1}{N}\sum_i^N\left\{3/2\left(\vec e_z\cdot\vec u_i\right)^2-1/2\right\}.
\end{equation}
 If $S=1$, the particles are perfectly aligned with the $z$-axis and if $S=-1/2$, particles are all perpendicular to the $z$-axis. The order parameter is visualised for two choices of the parameter $\hat{d}$ in Figure \ref{order_param}. We expect the order parameter to be $S(t)=1$ throughout the simulation, but numerically this only holds up until some point in time when the alignment is broken for some particles and $S(t)$ decreases. With varying choices of $\hat{d}$ and $\Delta t$, the same qualitative behaviour can be noted for the order parameter; The particles are initially ordered symmetrically and the background flow is symmetric, then, particles first start to come in contact in the vertical direction and later both vertically and horizontally. Due to contact forces not being perfectly symmetric, the alignment is broken.
 This divergence of the order-parameter however happens after a large number of time-steps for all ($\hat{d},\Delta t$), from which we can conclude that the contact handling is robust. 
\begin{table}[h!]
	\centering
	\begin{tabular}{c|c|c|c|c}
		 \shortstack[c]{Time-step $\Delta t$ as  \\ fraction of $\Delta t^* = 0.02$ \\ \quad} & \shortstack[c]{\# contact time-steps\\ $\hat{d} = 0.01R_{\text{rod}}$ \\ $T_{\text{end}} = 9$ ($T_S = 4.36$) } & \shortstack[c]{\# contact time-steps\\$\hat{d} = 0.1R_{\text{rod}}$\\ $T_{\text{end}}$ ($T_S$)} & \shortstack[c]{total \# steps\\ \vspace{2.5ex} \\ $T_{\text{end}}$ ($T_S$)} \\ \hline
		$\Delta t^*$ & 410 (178) & 412 (180) & 450 (218)\\
		$\Delta t^*/2$ &  819 (357) & 824 (362) & 900 (436) \\
		$\Delta t^*/4$ &  1639 (715) & 1647 (723) & 1800 (872) \\
	\end{tabular}
\caption{Rods in a biaxial compression flow: Number of time-steps where the contact algorithm is triggered for the different time-step sizes $\Delta t$, together with the total number of steps taken, with $T_S$ the time up until which the order parameter $S$ is displayed in Figure \ref{order_param} and $T_{\text{end}}$ the time when particles start to diverge along the $x$-axis (where the background flow is diverging).}
\label{tab:dt}
\end{table}

\clearpage
\subsection{Boomerangs}
\subsubsection{Geometric considerations}
We consider boomerang particles of the type displayed in Figure \ref{boomerang}. Such a boomerang is constructed by dividing a rod of aspect ratio $L_{\text{rod}}/R_{\text{rod}}=8$ and radius $R_{\text{rod}}=0.5$ as described in Section \ref{rod_type} into two equal pieces, each with a cylindrical part and a semi-spherical cap at one end. One of these half-rods is placed vertically and one placed horisontally so that the corners of the cut area of the two pieces touch. A circle is rotated around this point so that the two rod pieces are joined into one particle. The center coordinate $\vec x$ is defined to be the mass center of the particle.

For the barrier energy and the direction of contact forces and torques in $\vecmc D$, the contact distance $d^{\text{s}}$, based on surface-node-to-surface-node distances, is used, as the particle geometry is non-convex. For this purpose, the center curve of each particle is considered, given by two line segments and a quarter of a circle joining the two segments. For each point on the discretised surface of particle 1, we may easily compute the shortest distance to the center curve of particle 2, subtracting one particle radius to give the shortest distance between the surfaces. The corresponding source/target pair of surface points is added to the list of surface nodes used to construct the matrix $\vecmc D$ in \eqref{Ddef_C}. The procedure is repeated with reversed numbering of the two particles in the pair. 
\subsubsection{A random suspension of boomerangs}
\begin{figure}[h!]
	\centering
	\begin{subfigure}[b!]{0.4\textwidth}
		\centering
	\includegraphics[trim = {1.8cm 0.7cm 2cm 1.2cm},clip,width=1\textwidth]{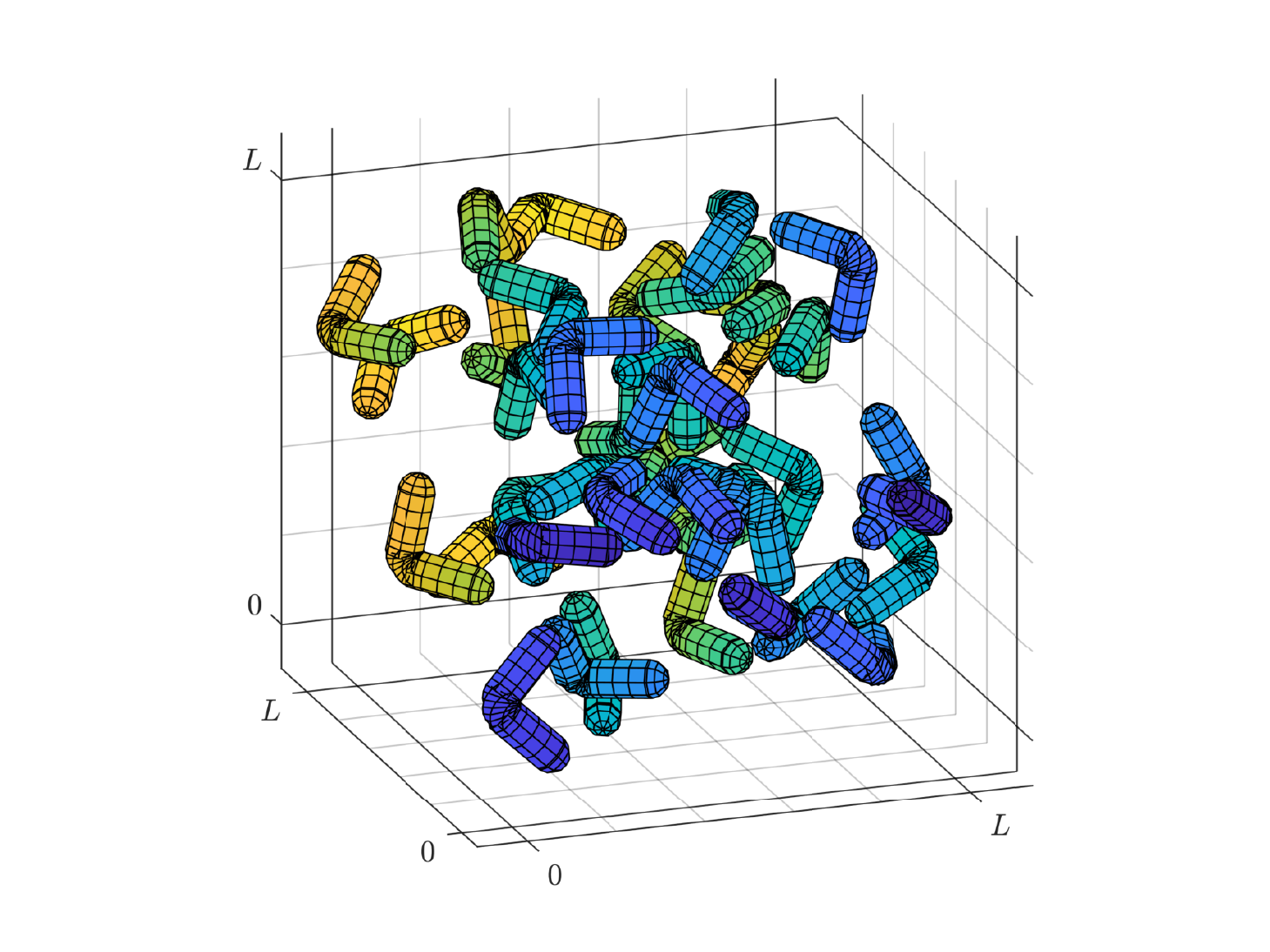}
	\caption{An example configuration, with colors indicating depth in the suspension.}
	\label{boomerangs}
	\end{subfigure}~~
	\begin{subfigure}[b!]{0.28\textwidth}
	\centering
	\includegraphics[trim = {1.5cm 18cm 13cm 3.3cm},clip,width=1\textwidth]{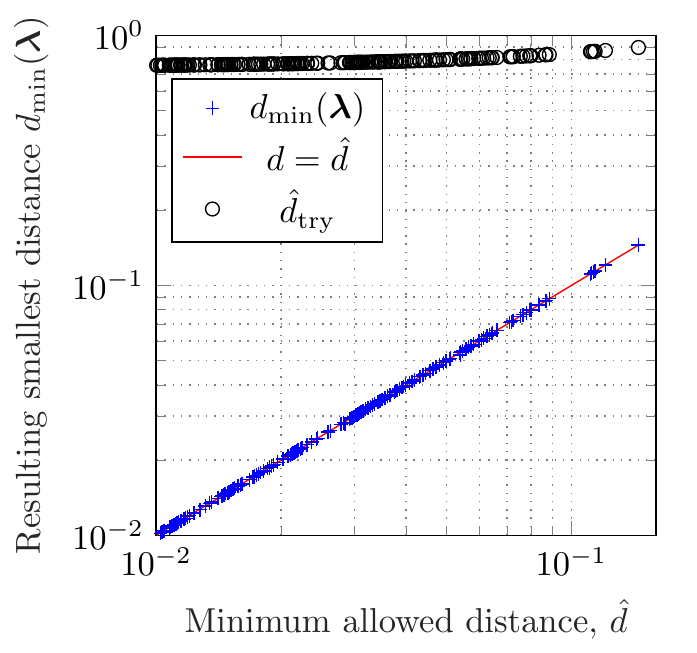}
	\caption{The minimum distances upon applying contact forces all respect the minimum allowed distance $\hat{d}$. For each configuration, the corresponding $\hat{d}_{\text{try}}$ is also displayed, determining the $N_c$ particle pairs flagged to be part of the contact force optimisation.}
	\label{smallest_boom}
\end{subfigure}~~
\begin{subfigure}[b!]{0.31\textwidth}
	\centering
	\includegraphics[trim = {1.5cm 17.8cm 11.5cm 3.3cm},clip,width=1.12\textwidth]{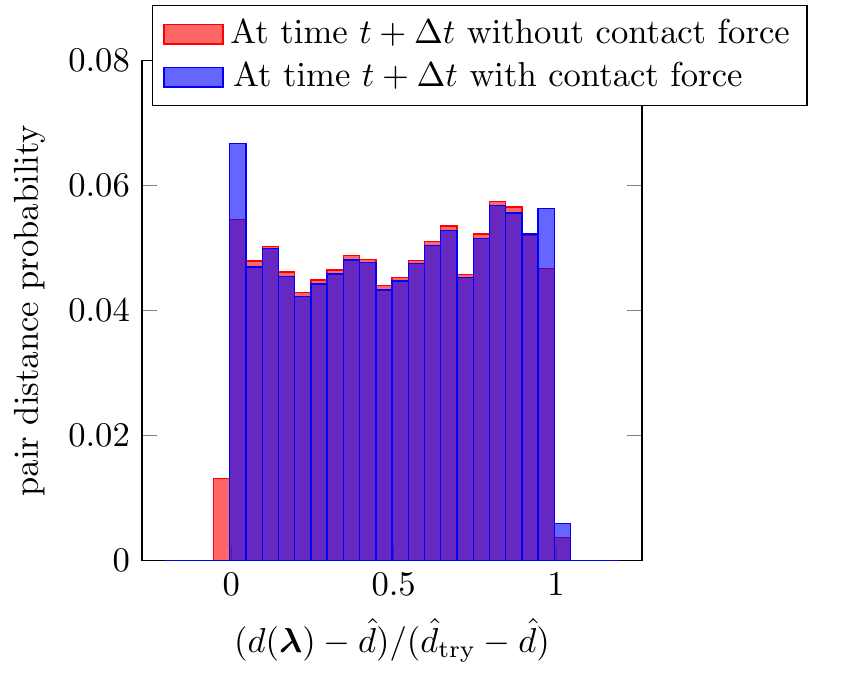}
	\caption{Inter-particle distances for the $N_c$ contact pairs before and after applying contact forces, with statistics collected from all 200 configurations. Note that $\hat{d}_{\text{try}}$ is large relative to $\hat{d}$, see (b).}
	\label{hist_boom}
\end{subfigure}
	\caption{Contact avoidance for 200 random configuration of 40 boomerangs.}
	\label{random_boom}
\end{figure}

Systems of 40 randomly positioned boomerangs in a cube of length $L = 12$ are considered, as exemplified in Figure \ref{boomerangs}. The boomerangs are generated one at the time with a uniformly sampled center coordinate $\vec x$ and quaternion $\vec q$ such that the distance to any other particle is larger than $\delta = 10^{-2}$.  We perform the same type of test as for the spheres in Section \ref{sphere_sec}: For each generated configuration of 40 boomerangs, the minimum separation distance is computed and $\hat{d}$ is chosen to be this distance. The boomerangs are then assigned forces and torques randomly sampled from a sphere with $\|\vecmc F_{\text{ext}}\|=100$ so that in a trial time-step with $\Delta t = 0.01$, some particles violate the constraints for surface-node-to-surface-node distances. For these particle pairs, correcting contact forces are computed. Among all the particle pairs flagged to potentially be assigned a contact force, the smallest distance obtained with contact forces at time $t+\Delta t$ is reported vs $\hat{d}$ in Figure \ref{smallest_boom} and all distances for the flagged pairs  are reported before and after the correction with contact forces in Figure \ref{hist_boom}. It can here be concluded that contact forces do not push particles apart unnecessarily far (note that the parameter $\hat{d}_{\text{try}}$ is chosen large relative to $\hat{d}$ not to miss any collisions). Despite the non-convex particle shape and the risk of particles getting stuck in locked configurations, $\hat{d}$ can be maintained for all the boomerangs in all of the 200 configurations. 

\section{A discussion on the computational cost}
Despite the increased cost in any time-step with contacts, a lot can be gained in terms of computational cost by applying the contact avoiding algorithm presented in this paper, as a much larger time-step can be considered for dynamical simulations than what would be allowed in a simulation without contact forces -- all time-step sizes used for the numerical experiments in this work are excessively large, but the contact algorithm is still robust. 

The main cost of finding optimal contact force magnitudes is the need for determining the matrix vector product $\vecmc M\vecmc D$ in the computation of the gradient of the barrier energy with respect to the vector of contact force magnittudes, which is needed in the interior point method used for solving the optimisation problem in \eqref{minnorm}. The matrix-matrix product $\vecmc M\vecmc D$ has to be computed and stored once per time-step where contact occurs, as $\vecmc M$ depends on $\vecmc Q_t$ and $\vecmc D$ depends on $\vecmc Q_{t+\Delta t}^*$ (no dependence on $\vec\lambda$). For the multiblob method, the mobility matrix can easily be computed explicitly if the total number of particles is small. For larger particle systems, evaluating $\vecmc M\vecmc D$ would amount to solving $N_c$ Stokes mobility problems, the cost of which is determined by $\hat{d}_{\text{try}}$ that will set $N_c$, which of course also depends on the particle density in the system. One Stokes mobility problem also has to be solved for every evaluation of the action of the contact forces, $\vecmc M\vecmc D\vec\lambda$, that enters in the barrier energy. The number of such solves depends on the number of iterations to solve the optimisation problem \eqref{minnorm} with the interior-point method, which, in turn, depends on the particle type, the time-step size and a few hyperparameters as discussed in Section \ref{sol_strat}: Choices that need to be made is e.g.~how to regularise the barrier function, i.e. how to handle negative distances between boundaries, and what stopping criteria to pick when solving for minimum contact force magnitudes in \eqref{minnorm}. From numerical experiments in this paper, it is expected that the number of mobility solves $\vecmc M\vecmc D\vec\lambda$ at least is in the range 10-20. The number of iterations is larger if the stopping criteria is very strict and smaller if the stopping criteria is less strict, which could be allowed by choosing a slightly larger buffer region $\hat{d}$ than required for accuracy in $\vecmc M$. We may compare the number of iterations to what is reported in the literature for the LCP technique in the work of Yan and collaborators, but at the same time emphasise that the iteration count will vary depending on the setting: Yan et al.~report $\approx 20$ extra Stokes solves per time-step for spheres in \cite{Yan2020} and 5-10 iterations for dilute suspensions of spherocylinders in \cite{Yan2019}, but $\mathcal O(1000)$ close to the  random-close-packing limit. 

The first-order interior-point method combined with the constrained minisation problem considered in this work also has benefits over second order methods, e.g.~based on Newton's method as applied to the non-constrained IPC-formulation \cite{Li2020}. In contrast to an IPC-formulation, we are not dependent on second order derivatives of the distances with respect to the particle coordinates and also avoid solving large linear systems -- other than the Stokes mobility problem -- while iteratively updating the solution vector in the optimisation problem. The matrix in the linear system in an IPC-formulation of the problem would be a function of $\vecmc M$, but also depend on the gradient of external forces and torques and on the Hessian of the distances  with respect to both particle center coordinates and quaternions. The structure of this matrix is not trivially investigated for the general case a priori and the matrix has to be projected onto the cone of nonnegative matrices for Newton's method to converge. Moreover, fast matrix-vector techniques do not apply as for the Stokes mobility problem (applying $\vec M\vecmc F$, for some vector $\vecmc F$) \cite{USABIAGA2016}. Because of these difficulties, the current formulation coupled to a first order method, such as the interior-point method, is especially beneficial in very large particle systems,  where we neither want to compute $\vecmc M$ explicitly nor want to repeatedly solve a large linear system that is a non-trivial function of $\vecmc M$\footnote{An option for larger particle systems could potentially be to solve the problem by employing the sparse nonlinear optimiser provided in the SNOPT package \cite{gill2005snopt}. Care however has to be taken for how second order derivatives are computed.}.
\section{Conclusions}
We have presented an optimisation procedure to guarantee non-overlapping configurations of 3D particles in an unbounded Stokesian fluid. The method is based on a barrier formulation where non-overlapping constraints are rewritten as a barrier energy, constrained to be zero for non-overlapping configurations. Numerical examples are provided for spheres, rod-like particles with semi-spherical caps and boomerangs. The numerical examples show the performance of the contact force algorithm for dense suspensions of particles deliberately pushed together by external forces or a background flow. In all examples, the highlighted property of the proposed method is its ability to keep a set minimum separation distance $\hat{d}$. Discrete complementarity is obtained at the trial time-step between $\vec\lambda$ and the modified barrier energy (with parameter $\hat{d}_{\text{try}}$). We have numerically shown that the method is robust for different choices of $\hat{d}$, particle shapes and flow scenarios. The magnitudes of external forces, background flows and time-step sizes are in this work deliberately chosen to force particles to come too close to each other and in a realistic simulation, contact will occur less frequently and $\hat{d}$ is supposed to be small.  Strengths of the method is its simple formulation, its ability of handling non-convex particles, the no-collision guarantee of the formulation and the minimum impact on the system by contact forces as these are balanced and their magnitudes are minimised.  By solving for force magnitudes explicitly, as we do here, a penalisation parameter can be omitted, which otherwise has to be carefully tuned. Moreover, the size of the optimisation problem is much more moderate (at least for moderatly crowded systems) than if all particle coordinates are solved for implicitly, as in Incremental Potential Contact (IPC) \cite{Li2020}.

For future work we identify two directions: 
\begin{enumerate}
	\item Approximation of the action of the mobility matrix: There is a global coupling between all particles in the suspension in the mobility matrix, $\vecmc M$. Motivated by the fact that close interactions are dominating, one option is to build an approximation to the action of $\vecmc M$ for matrix-vector multiplies $\vecmc M\vecmc F$ constructed only from all pairs of particles in contact, instead of applying the global mobility matrix. An issue is however that a contact force implying no-collision determined with such an approximation does not necessarily have to lead to a non-overlapping configuration with $\vecmc M$. A workaround could be to introduce a small buffer for the minimum allowed distance so that $\hat{d}\to\hat{d}(1+\delta)$ or to use the approximated solution vector for contact force magnitudes as an initial guess for iterations with the full mobility matrix. Along similar lines, separated clusters of particles could be considered to compute contact forces locally.
	\item A continuous contact force potential: In this work, a discrete version of the contact force potential is considered, computed from all points on the particle surfaces sufficiently close to each other. One could also consider a continuously defined potential for two particles in contact, where the barrier energy at the trial time-step is expressed in terms of integrals over the particle surfaces close to contact.  The asymmetry of the contact forces noted in the experiment with a biaxial compression flow in Section \ref{biaxial} has two explanations: To start with, the discrete surface grid on the particles results in different barrier energies for seemingly similar relative rotations and particle-particle distances for different colliding pairs -- depending on the rotations of the particles around their own axis, a different number of grid nodes may be flagged for collision. Furthermore, the shortest particle-particle distance $d^{\text{p}}$ is sensitive to slight differences in orientation when particles are close to parallel.  One benefit of a continuous contact force potential could be to better preserve symmetries. We leave this approach for future work starting in 2D.
\end{enumerate}

We end with a note on the optimisation algorithm: In this work, the inbuilt optimisation procedure \texttt{fmincon} in \texttt{Matlab} has been used to solve the optimisation problem for the contact force magnitude vector $\vec\lambda$, employing a highly tuned interior point method. The fact that a standard solver in \textsc{Matlab} can be used off-the-shelf is a strength of the method. We are satisfied with a possibly \emph{local} optimum as long as a zero barrier energy is reached, implying sufficiently separated particles.


\section*{Acknowledgements}
The authors thank Anders Forsgren for discussions on the choice of optimisation algorithm, Georg Stadler for discussing alternative ways of setting up the non-overlap constraints
and Mattias Sandberg for constructive input on the numerical examples provided in this work. We acknowledge the support from the Swedish Research Council: grant no.~2019-05206 and the research environment grant INTERFACE (biomaterials), no.~2016-06119.

\addcontentsline{toc}{section}{References}
\bibliographystyle{myIEEEtran} 
\bibliography{contact_refs}

\begin{thebibliography}{10}
\providecommand{\url}[1]{#1}
\csname url@samestyle\endcsname
\providecommand{\newblock}{\relax}
\providecommand{\bibinfo}[2]{#2}
\providecommand{\BIBentrySTDinterwordspacing}{\spaceskip=0pt\relax}
\providecommand{\BIBentryALTinterwordstretchfactor}{4}
\providecommand{\BIBentryALTinterwordspacing}{\spaceskip=\fontdimen2\font plus
\BIBentryALTinterwordstretchfactor\fontdimen3\font minus
  \fontdimen4\font\relax}
\providecommand{\BIBforeignlanguage}[2]{{%
\expandafter\ifx\csname l@#1\endcsname\relax
\typeout{** WARNING: IEEEtran.bst: No hyphenation pattern has been}%
\typeout{** loaded for the language `#1'. Using the pattern for}%
\typeout{** the default language instead.}%
\else
\language=\csname l@#1\endcsname
\fi
#2}}
\providecommand{\BIBdecl}{\relax}
\BIBdecl

\bibitem{USABIAGA2016}
F.~B. Usabiaga, B.~Kallemov, B.~Delmotte, A.~P.~S. Bhalla, A.~Donev, and B.~E.
  Griffith, ``{Hydrodynamics of Suspensions of Passive and Active Rigid
  Particles : A Rigid Multiblob Approach},'' \emph{Comm. App. Math. Comp.
  Sci.}, vol.~11, no.~2,  2016 doi:
  \href{http://dx.doi.org/10.2140/camcos.2016.11.217}{10.2140/camcos.2016.11.217}

\bibitem{Broms2022}
A.~Broms, M.~Sandberg, and A.-K. Tornberg, ``A locally corrected multiblob
  method with hydrodynamically matched grids for the stokes mobility problem,''
   2022 doi:
  \href{http://dx.doi.org/10.48550/ARXIV.2207.11210}{10.48550/ARXIV.2207.11210}

\bibitem{Delong2015}
\BIBentryALTinterwordspacing
S.~Delong, F.~{Balboa Usabiaga}, and A.~Donev, ``{Brownian dynamics of confined
  rigid bodies},'' \emph{J. Chem. Phys}, vol. 143, no.~14,  2015 doi:
  \href{http://dx.doi.org/10.1063/1.4932062}{10.1063/1.4932062}
\BIBentrySTDinterwordspacing

\bibitem{Sprinkle2017}
\BIBentryALTinterwordspacing
B.~Sprinkle, F.~{Balboa Usabiaga}, N.~A. Patankar, and A.~Donev, ``{Large scale
  Brownian dynamics of confined suspensions of rigid particles},'' \emph{J.
  Chem. Phys}, vol. 147, no.~24,  2017 doi:
  \href{http://dx.doi.org/10.1063/1.5003833}{10.1063/1.5003833}
\BIBentrySTDinterwordspacing

\bibitem{Brosseau2019}
\BIBentryALTinterwordspacing
Q.~Brosseau, F.~B. Usabiaga, E.~Lushi, Y.~Wu, L.~Ristroph, J.~Zhang, M.~Ward,
  and M.~J. Shelley, ``{Relating rheotaxis and hydrodynamic actuation using
  asymmetric gold-platinum phoretic rods},'' \emph{Phys. Rev. Lett}, vol. 123,
  no.~17, p. 178004,  2019 doi:
  \href{http://dx.doi.org/10.1103/PhysRevLett.123.178004}{10.1103/PhysRevLett.123.178004}
\BIBentrySTDinterwordspacing

\bibitem{Brosseau2021}
Q.~Brosseau, F.~B. Usabiaga, E.~Lushi, Y.~Wu, L.~Ristroph, M.~D. Ward, M.~J.
  Shelley, and J.~Zhang, ``{Metallic microswimmers driven up the wall by
  gravity},'' \emph{Soft Matter}, vol.~17, no.~27, pp. 6597--6602,  2021 doi:
  \href{http://dx.doi.org/10.1039/d1sm00554e}{10.1039/d1sm00554e}

\bibitem{Fiore2019}
A.~M. Fiore and J.~W. Swan, ``{Fast Stokesian dynamics},'' \emph{J. Fluid
  Mech.}, vol. 878, pp. 544--597,  2019 doi:
  \href{http://dx.doi.org/10.1017/jfm.2019.640}{10.1017/jfm.2019.640}

\bibitem{AfKlinteberg2014}
L.~{af Klinteberg} and A.~K. Tornberg, ``{Fast Ewald summation for Stokesian
  particle suspensions},'' \emph{Int. J. Numer. Methods Fluids}, vol.~76,
  no.~10, pp. 669--698,  2014 doi:
  \href{http://dx.doi.org/10.1002/fld.3953}{10.1002/fld.3953}

\bibitem{Berne1972}
B.~J. Berne and P.~Pechukas, ``{Gaussian model potentials for molecular
  interactions},'' \emph{J. Chem. Phys}, vol.~56, no.~8, pp. 4195--4205,  1972
  doi: \href{http://dx.doi.org/10.1063/1.1677837}{10.1063/1.1677837}

\bibitem{Varga2006}
S.~Varga and G.~Jackson, ``{Study of the pitch of fluids of electrostatically
  chiral anisotropic molecules: Mean-field theory and simulation},'' \emph{Mol.
  Phys.}, vol. 104, no. 22-24, pp. 3681--3691,  2006 doi:
  \href{http://dx.doi.org/10.1080/00268970601058556}{10.1080/00268970601058556}

\bibitem{Tao2005}
Y.~G. Tao, W.~K. {Den Otter}, J.~T. Padding, J.~K. Dhont, and W.~J. Briels,
  ``{Brownian dynamics simulations of the self- and collective rotational
  diffusion coefficients of rigid long thin rods},'' \emph{J. Chem. Phys}, vol.
  122, no.~24,  2005 doi:
  \href{http://dx.doi.org/10.1063/1.1940031}{10.1063/1.1940031}

\bibitem{Baraff1993}
D.~Baraff, ``{Issues in computing contact forces for non-penetrating rigid
  bodies},'' \emph{Algorithmica}, vol.~10, no. 2-4, pp. 292--352,  1993 doi:
  \href{http://dx.doi.org/10.1007/BF01891843}{10.1007/BF01891843}

\bibitem{Anitescu1996}
Anitescu, Cremer, and Potra, ``{Formulating 3D Contact Dynamics Problems},''
  \emph{Mechanics of Structures and Machines}, vol.~24, no.~4, pp. 405--437,
  1996 doi:
  \href{http://dx.doi.org/10.1080/08905459608905271}{10.1080/08905459608905271}

\bibitem{Tasora2008}
A.~Tasora and M.~Anitescu, ``{A Fast NCP Solver for Large Rigid-Body Problems
  with Contacts, Friction, and Joints},'' \emph{Multibody Dynamics}, pp.
  45--55,  2008 doi:
  \href{http://dx.doi.org/10.1007/978-1-4020-8829-2\textunderscore3}{10.1007/978-1-4020-8829-2\textunderscore3}

\bibitem{Tasora2008a}
A.~Tasora, D.~Negrut, and M.~Anitescu, ``{Large-scale parallel multi-body
  dynamics with frictional contact on the graphical processing unit},''
  \emph{Proc. Inst. Mech. Eng., Part K: J. Multi-Body Dyn}, vol. 222, no.~4,
  pp. 315--326,  2008 doi:
  \href{http://dx.doi.org/10.1243/14644193JMBD154}{10.1243/14644193JMBD154}

\bibitem{Tur2009}
\BIBentryALTinterwordspacing
M.~Tur, F.~J. Fuenmayor, and P.~Wriggers, ``{A mortar-based frictional contact
  formulation for large deformations using Lagrange multipliers},''
  \emph{Comput. Methods Appl. Mech. Eng.}, vol. 198, no. 37-40, pp. 2860--2873,
   2009 doi:
  \href{http://dx.doi.org/10.1016/j.cma.2009.04.007}{10.1016/j.cma.2009.04.007}
\BIBentrySTDinterwordspacing

\bibitem{Wriggers2006}
P.~Wriggers, \emph{{Computational contact mechanics, second ed., Springer}},
  2006.  ISBN 9783540326083 doi:
  \href{http://dx.doi.org/10.1007/978-3-540-32609-0}{10.1007/978-3-540-32609-0}

\bibitem{Snyder1995}
J.~M. Snyder, ``{Interactive tool for placing curved surfaces without
  interpenetration},'' \emph{Proceedings of the ACM SIGGRAPH Conference on
  Computer Graphics}, pp. 209--218,  1995 doi:
  \href{http://dx.doi.org/10.1145/218380.218444}{10.1145/218380.218444}

\bibitem{Harmon2011}
\BIBentryALTinterwordspacing
D.~Harmon, D.~Panozzo, O.~Sorkine, and D.~Zorin, ``{Interference-aware
  geometric modeling},'' \emph{ACM Trans. Graph.}, vol.~30, no.~6, p.~1,  2011
  doi:
  \href{http://dx.doi.org/10.1145/2070781.2024171}{10.1145/2070781.2024171}
\BIBentrySTDinterwordspacing

\bibitem{Li2020}
M.~Li, Z.~Ferguson, T.~Schneider, T.~Langlois, D.~Zorin, D.~Panozzo, C.~Jiang,
  and D.~M. Kaufman, ``{Incremental Potential Contact: Intersection- and
  Inversion-free, Large-Deformation Dynamics},'' \emph{ACM Trans. Graph.},
  vol.~39, no.~4,  2020 doi:
  \href{http://dx.doi.org/10.1145/3386569.3392425}{10.1145/3386569.3392425}

\bibitem{Ferguson2021}
Z.~Ferguson, M.~Li, T.~Schneider, F.~Gil-Ureta, T.~Langlois, C.~Jiang,
  D.~Zorin, D.~M. Kaufman, and D.~Panozzo, ``{Intersection-free rigid body
  dynamics},'' \emph{ACM Trans. Graph.}, vol.~40, no.~4, pp. 1--16,  2021 doi:
  \href{http://dx.doi.org/10.1145/3476576.3476773}{10.1145/3476576.3476773}

\bibitem{Li2021}
M.~Li, D.~M. Kaufman, and C.~Jiang, ``{Codimensional incremental potential
  contact},'' \emph{ACM Trans. Graph.}, vol.~40, no.~4,  2021 doi:
  \href{http://dx.doi.org/10.1145/3450626.3459767}{10.1145/3450626.3459767}

\bibitem{Yamamoto1993}
S.~Yamamoto and T.~Matsuoka, ``{A method for dynamic simulation of rigid and
  flexible fibers in a flow field},'' \emph{J. Chem. Phys}, vol.~98, no.~1, pp.
  644--650,  1993 doi:
  \href{http://dx.doi.org/10.1063/1.464607}{10.1063/1.464607}

\bibitem{Yamamoto1995}
------, ``{Dynamic simulation of fiber suspensions in shear flow},'' \emph{J.
  Chem. Phys}, vol. 102, no.~5, pp. 2254--2260,  1995 doi:
  \href{http://dx.doi.org/10.1063/1.468746}{10.1063/1.468746}

\bibitem{Das2013}
D.~Das and D.~Saintillan, ``{Electrohydrodynamic interaction of spherical
  particles under Quincke rotation},'' \emph{Phys. Rev. E}, vol.~87, no.~4, pp.
  1--14,  2013 doi:
  \href{http://dx.doi.org/10.1103/PhysRevE.87.043014}{10.1103/PhysRevE.87.043014}

\bibitem{Delmotte2015b}
\BIBentryALTinterwordspacing
B.~Delmotte, E.~Climent, and F.~Plourabou{\'{e}}, ``{A general formulation of
  Bead Models applied to flexible fibers and active filaments at low Reynolds
  number},'' \emph{J. Comput. Phys}, vol. 286, pp. 14--37,  2015 doi:
  \href{http://dx.doi.org/10.1016/j.jcp.2015.01.026}{10.1016/j.jcp.2015.01.026}
\BIBentrySTDinterwordspacing

\bibitem{Corona2017}
\BIBentryALTinterwordspacing
E.~Corona, L.~Greengard, M.~Rachh, and S.~Veerapaneni, ``{An integral equation
  formulation for rigid bodies in Stokes flow in three dimensions},'' \emph{J.
  Comput. Phys}, vol. 332, pp. 504--519,  2017 doi:
  \href{http://dx.doi.org/10.1016/j.jcp.2016.12.018}{10.1016/j.jcp.2016.12.018}
\BIBentrySTDinterwordspacing

\bibitem{Lu2019}
L.~Lu, ``{Parallel contact-aware algorithms for large-scale direct blood flow
  simulations},'' Ph.D. dissertation, Courant Institute of Mathematical
  Sciences, New York University, 2019.

\bibitem{Yan2019}
W.~Yan, H.~Zhang, and M.~J. Shelley, ``{Computing collision stress in
  assemblies of active spherocylinders: Applications of a fast and generic
  geometric method},'' \emph{J. Chem. Phys}, vol. 150, no.~6,  2019 doi:
  \href{http://dx.doi.org/10.1063/1.5080433}{10.1063/1.5080433}

\bibitem{Yan2020}
W.~Yan, E.~Corona, D.~Malhotra, S.~Veerapaneni, and M.~Shelley, ``{A scalable
  computational platform for particulate Stokes suspensions},'' \emph{J.
  Comput. Phys}, vol. 416, p. 109524,  2020 doi:
  \href{http://dx.doi.org/10.1016/j.jcp.2020.109524}{10.1016/j.jcp.2020.109524}

\bibitem{Bystricky2020}
L.~Bystricky, S.~Shanbhag, and B.~Quaife, ``{Stable and contact-free time
  stepping for dense rigid particle suspensions},'' \emph{Int. J. Numer.
  Methods Fluids}, vol.~92, no.~2, pp. 94--113,  2020 doi:
  \href{http://dx.doi.org/10.1002/fld.4774}{10.1002/fld.4774}

\bibitem{Yan2022}
W.~Yan, S.~Ansari, A.~Lamson, M.~A. Glaser, R.~Blackwell, M.~D. Betterton, and
  M.~Shelley, ``{Toward the cellular- ­ scale simulation of motor- ­ driven
  cytoskeletal assemblies},'' \emph{eLife}, no.~11, p. e74160,  2022 doi:
  \href{http://dx.doi.org/10.7554/eLife.74160}{10.7554/eLife.74160}

\bibitem{Kohl2021}
\BIBentryALTinterwordspacing
R.~Kohl, E.~Corona, V.~Cheruvu, and S.~Veerapaneni, ``{Fast and accurate
  solvers for simulating Janus particle suspensions in Stokes flow},'' pp.
  1--24,  2021 doi:
  \href{http://dx.doi.org/10.48550/arXiv.2104.14068}{10.48550/arXiv.2104.14068}
\BIBentrySTDinterwordspacing

\bibitem{Lu2019a}
L.~Lu, M.~J. Morse, A.~Rahimian, G.~Stadler, and D.~Zorin, ``{Scalable
  simulation of realistic volume fraction red blood cell flows through vascular
  networks},'' \emph{Int. Conf. High Perform. Comput. Netw. Storage Anal.},
  2019 doi:
  \href{http://dx.doi.org/10.1145/3295500.3356203}{10.1145/3295500.3356203}

\bibitem{Bystricky2018t}
L.~Bystricky, ``{Contact-free Simulations of Rigid Particle Suspensions Using
  Boundary Integral Equations},'' Ph.D. dissertation, Florida State University.
  ISBN 9780438447769 2018.

\bibitem{Lu2017}
\BIBentryALTinterwordspacing
L.~Lu, A.~Rahimian, and D.~Zorin, ``{Contact-aware simulations of particulate
  Stokesian suspensions},'' \emph{J. Comput. Phys}, vol. 347, pp. 160--182,
  2017 doi:
  \href{http://dx.doi.org/10.1016/j.jcp.2017.06.039}{10.1016/j.jcp.2017.06.039}
\BIBentrySTDinterwordspacing

\bibitem{Boyd2009}
S.~Boyd and L.~Vandenberghe, \emph{{Convex optimization}},  2009 doi:
  \href{http://dx.doi.org/10.1142/9789814412520\textunderscore0010}{10.1142/9789814412520\textunderscore0010}

\bibitem{LUMELSKY1985}
V.~J. Lumelsky, ``On fast computation of distance between line segments,''
  \emph{Inf Process Lett}, vol.~21, no.~2, pp. 55--61,  1985 doi:
  \href{http://dx.doi.org/10.1016/0020-0190(85)90032-8}{10.1016/0020-0190(85)90032-8}

\bibitem{Ondrej2015}
Ondrej, ``Fast shortest distance between two line segments (in n dimensions),''
  MATLAB Central File Exchange, Retrieved May 4, 2022,
  \url{https://www.mathworks.com/matlabcentral/fileexchange/49502-fast-shortest-distance-between-two-line-segments-in-n-dimensions}.

\bibitem{Fischer1992}
A.~Fischer, ``{A special newton-type optimization method},''
  \emph{Optimization}, vol.~24, no. 3-4, pp. 269--284,  1992 doi:
  \href{http://dx.doi.org/10.1080/02331939208843795}{10.1080/02331939208843795}

\bibitem{Dai2005}
Y.~H. Dai and R.~Fletcher, ``{Projected Barzilai-Borwein methods for
  large-scale box-constrained quadratic programming},'' \emph{Numerische
  Mathematik}, vol. 100, no.~1, pp. 21--47,  2005 doi:
  \href{http://dx.doi.org/10.1007/s00211-004-0569-y}{10.1007/s00211-004-0569-y}

\bibitem{Fletcher2005}
R.~Fletcher, ``On the barzilai-borwein method,'' in \emph{Optimization and
  Control with Applications}, L.~Qi, K.~Teo, and X.~Yang, Eds.\hskip 1em plus
  0.5em minus 0.4em\relax  Boston, MA: Springer US, 2005 doi:
  \href{http://dx.doi.org/10.1007/b104943}{10.1007/b104943}. ISBN
  978-0-387-24255-2 pp. 235--256.

\bibitem{Doi2019}
H.~Doi, K.~Z. Takahashi, K.~Tagashira, J.~ichi Fukuda, and T.~Aoyagi,
  ``{Machine learning-aided analysis for complex local structure of liquid
  crystal polymers},'' \emph{Sci. Rep.}, vol.~9, no.~1, pp. 1--12,  2019 doi:
  \href{http://dx.doi.org/10.1038/s41598-019-51238-1}{10.1038/s41598-019-51238-1}

\bibitem{gill2005snopt}
P.~E. Gill, W.~Murray, and M.~A. Saunders, ``Snopt: An sqp algorithm for
  large-scale constrained optimization,'' \emph{SIAM Rev.}, vol.~47, no.~1, pp.
  99--131,  2005 doi:
  \href{http://dx.doi.org/10.1137/S0036144504446096}{10.1137/S0036144504446096}

\end{thebibliography}

\end{document}